\documentclass[aps,twocolumn,
floats,prl,nofootinbib,superscriptaddress,10pt]{revtex4-1}
\usepackage{float}
\usepackage[utf8]{inputenc}

\usepackage{graphicx,amsmath,amsfonts,amssymb,slashed}
\usepackage{bbold,wasysym}
\usepackage{enumitem}

\usepackage{amsmath,amsfonts,bbm,euscript,hyperref}

\usepackage{xcolor}
\usepackage{color}
\usepackage{hyperref}

\hypersetup{colorlinks, citecolor=blue, linkcolor=red, urlcolor=blue}

\usepackage{graphicx,capt-of}

\newcounter{subfloat}

\newcounter{subfloat2}

\newcommand{\nco}{\newcommand}
\nco{\beq}{\begin{equation}} \nco{\eeq}{\end{equation}}
\nco{\beqa}{\begin{eqnarray}} \nco{\eeqa}{\end{eqnarray}}
\def\be{\begin{equation}}
\def\ee{\end{equation}}    
\def\baray{\begin{eqnarray}}
\def\earay{\end{eqnarray}}
\def\rt{\tilde r}

\nco{\lra}{\leftrightarrow}

\nco{\sss}{\scriptscriptstyle} \nco{\dphi}{\varphi}
\nco{\lsim}{\mbox{\raisebox{-.6ex}{~$\stackrel{<}{\sim}$~}}}
\nco{\gsim}{\mbox{\raisebox{-.6ex}{~$\stackrel{>}{\sim}$~}}}

\def\IK{\relax{\rm I\kern-.20em K}}
\def\IM{\relax{\rm I\kern-.20em M}}

\def\lsim{\mbox{\raisebox{-.6ex}{~$\stackrel{<}{\sim}$~}}}
\def\gsim{\mbox{\raisebox{-.6ex}{~$\stackrel{>}{\sim}$~}}}
\def\sss{\scriptscriptstyle}

\def\ga{\mathrel{\raise.3ex\hbox{$>$\kern-.75em\lower1ex\hbox{$\sim$}}}}
\def\la{\mathrel{\raise.3ex\hbox{$<$\kern-.75em\lower1ex\hbox{$\sim$}}}}

\usepackage{amsmath}
\usepackage{lipsum}

\begin{document}

\title{ Accretion, Evaporation and Superradiance Phase Diagram of  (Primordial) Black Holes  \\ and   $10^{-21}-10^{21}$  eV  Scalar, Vector, Tensor Fields
}
\author{Caner \"Unal} 
\email{cerul2870@gmail.com}
\affiliation{Department of Physics, Ben-Gurion University of the Negev, Be'er Sheva 84105, Israel}
\affiliation{Department of Physics, Middle East Technical University, 06800 Ankara, Turkey}
\affiliation{Feza Gursey Institute, Bogazici University, Kandilli, 34684, Istanbul, Turkey}

\begin{abstract} 
We obtain the accretion, evaporation and superradiance phase diagram of astrophysical and primordial black holes in the mass range $10^{-33}-10^{11} \, M_\odot $. This black hole mass range corresponds to production of $10^{-21} - 10^{21}$ eV particles for superradiance (bosons) and evaporation (bosons and fermions). Only accretion and superradiance processes are relevant for heavy black holes, on the other hand for light black holes of primordial origin, all three processes can be relevant.  We find that superradiance instability can happen even for black hole spin values as low as $10^{-9}-10^{-2}$. Since light black holes are very unstable to these perturbations and sensitive probes of bosonic particles, a single moderately spinning BH can probe 2-9 orders of magnitude of mass parameter space depending on the nature of the perturbations that are scalar (axion), vector (dark photon and/or photon with effective mass) and spin-2. If spinning black holes are observed and superradiance is not observed, possibly due to self-interactions, we find limits on the axion/scalar decay constant and energy density. We generalize these bounds for vector and spin-2 fields.
\end{abstract}

\maketitle
\noindent
{\bf Introduction.} Black holes (BHs) formed via astrophysical collapse have mass over 2.5 $M_\odot$ and can grow upto $10^{11}M_\odot$. BHs with any mass from $10^{-33}-10^{11}M_\odot$ can also come from cosmological primordial/early universe processes called as primordial black holes \cite{Green:2020jor,Carr:2020xqk,Carr:2020gox}  \footnote{
See  \cite{Sasaki:2016jop,Raidal:2017mfl,Ali-Haimoud:2017rtz} for merging PBH stochastic GW, and \cite{Aggarwal:2020olq,Franciolini:2022htd} for GW signatures of light BHs.}. Parameter space for primordial BHs is still open in such a way that they can make up nearly all dark matter in the mass range $10^{-15}-10^{-11} M_\odot$ \cite{Saito:2008jc,Garcia-Bellido:2017aan,Domcke:2017fix,Bartolo:2018rku,Cai:2018dig,Unal:2018yaa}, and  sub-percent fraction of dark matter in the range $10^{-6}-100 M_\odot$ \cite{Bird:2016dcv,Sasaki:2016jop,Clesse:2016vqa} \footnote{ Recent lensing data can be related to $10^{-6}-10^{-3} M_\odot$ BHs \cite{Niikura:2019kqi}.}.
Hence, light BHs can make up part of dark content. Even if they form a small fraction of the energy density of our universe, they might have important implications on inflationary stage, field content, primordial universe, and the evolution of our universe.

Black holes, independent from their formation mechanism, accrete, evaporate (both femrions and bosons) and superradiate (if there exist bosonic degrees of freedom whose mass is comparable to black hole horizon scale). We obtain the phase diagram of black hole phases that are i)evaporation, ii)accretion and iii)superradiance for scalar, vector and tensor fields in the mass-spin plane, in the black hole mass range $10^{-33}-10^{11} \; M_\odot$ ($\sim  1-10^{44} \; {\rm gr}$), corresponding to particle mass range $10^{-21}-10^{21}$ eV. 
We find the boundaries between these processes in the given mass range.
In result, we obtain the general phase diagram of black hole evolution in the presence of scalar, vector and tensor degrees of freedom (dof).  In this work, although we give results for all the mass regime, we especially focus on BHs lighter than solar mass, $(10^{-33}-1) \, M_\odot$ since superradiance has been worked out for the heavy black holes, ie. stellar mass BHs ($1-10^2 M_\odot$) \cite{SRcond,SRcond1,SRreview,Stott:2020gjj} 
and supermassive BHs ($10^6-10^{10} \, M_\odot$) \cite{SRcond,SRcond1,SRreview,Stott:2020gjj,Unal:2020jiy,Roy:2021uye,Chen:2022nbb,Chen:2021lvo}.  For black holes larger than $10^{-20}M_\odot$ only superradiance and accretion processes are relevant. On the other hand for black holes lighter than $10^{-20}M_\odot$ only superradiance and evaporation processes are relevant. Light primordial black holes can probe heavier particles via evaporation and superradiance.

 We derive the minimum BH spin required to have superradiance for scalar, vector and tensor perturbations as a function of BH mass, then we further include the effect of Hawking evaporation. We find that superradiance can happen even for very low spin values, ie. $a \sim 10^{-4}-10^{-2}$ for scalars, $a \sim 10^{-6}-10^{-3}$ for vectors, and $a \sim 10^{-9}-10^{-4}$ for spin-2 tensors.  Therefore, if perturbed by bosons, for example axions \cite{SRcond,SRcond1}, Higgs boson and Standard Model and Beyond Standard Model scalars/pseudoscalars, dark photons and photons with effective mass from plasma interactions \cite{Pani:2013hpa,Baryakhtar:2017ngi,Conlon:2017hhi,Dima:2020rzg,Cannizzaro:2020uap,Cannizzaro:2021zbp}, rotating light black holes deplete their spin rapidly independent of their initial formation spin \cite{Flores:2021tmc} \footnote{Small spin is expected with horizon size collapse in radiation domination \cite{Mirbabayi:2019uph,DeLuca:2019buf}, or large spin in matter era \cite{Harada:2017fjm}. BH spin can reach moderate values ${\tilde a} \sim 0.7$ via mergers, and for light BHs, accretion may not change the initial mass and spin \cite{DeLuca:2020bjf}.}. Since they are very unstable to these perturbations and sensitive probes of bosonic particles, a single moderately spinning BH can probe/cover 2-9 orders of magnitude scalar (axion), vector (dark photon and/or photon with effective mass) and spin-2 mass \footnote{See also 
~\cite{Hlozek:2014lca,Poulin:2018dzj,Lague:2021frh,Bozek:2014uqa,Kobayashi:2017jcf,Irsic:2017yje,Ferreira:2020fam,Armengaud:2017nkf,Rogers:2020ltq,Rogers:2020cup,Marsh:2018zyw,Bar:2021kti,Unal:2020jiy,SRcond,Tsukada:2018mbp,Tsukada:2020lgt,Yuan:2022bem,Unal:2022ooa,Blas:2019hxz,Armaleo:2019gil,LopezNacir:2018epg,Blas:2016ddr} 
for various multi-messenger probes. 
}.  Detection of superradiance would have remarkable implications for bosonic degrees of freedom (evaporation for both bosons and fermionic), but in the case that superradiance is not observed (spinning black holes are observed), say due to self-interactions  \cite{Baryakhtar:2020gao,Blas:2020kaa}, we find limits on the corresponding boson's axion decay constant  and energy density.

\vspace{0.2cm}
\noindent
{\bf BH Superradiance.} Spinning BHs can deplete their rotational energy into bosonic particles via superradiance \cite{Penrose:1969pc,Misner:1972kx,ZelDovich1972,Starobinsky:1973aij,Teukolsky:1974yv,Ternov:1978gq,Detweiler:1980uk}. The instability rate of black hole is different for distinct types of perturbations. When the Compton wavelength of the bosonic fields become comparable to the horizon size, they couple to the BH and can extract energy and angular momentum from it.  First condition is BH has larger angular velocity than the field
\begin{equation}
\mu_b < m \, \Omega_H \, ,
\label{spincond}
\end{equation}
$m$ being the azimuthal number and $\Omega_H (w_H) $ the angular speed (dimensionless angular speed), defined as
\begin{equation}
\Omega_H \equiv \frac{a}{2 r_g \left(1+\sqrt{1-a^2}\right)}  = \frac{1}{2r_g} \; w_H \, ,
\end{equation}
$r_g=G \, M_{BH}$ is the gravitational radius. We will suppress the subscript BH for the rest of the paper.

Besides \eqref{spincond}, one also requires instability rate is faster than accretion rate, or equivalently the characteristic time scale for BH accretion is longer than the instability time scale \cite{SRcond,SRcond1,SRreview}
\beq
 \tau_{Acc} > \tau_{SR} 
 \label{timecond}
\eeq
Accreting black holes build up mass and spin (this  usually depends on how chaotic accretion is). When BH has twice its mass via thin disk and smooth accretion, the spin reaches nearly maximum value \cite{refnovikovthorne}. Here we estimate the BH time scale starting from accretion rate as
\beq
{\dot M_{acc}} = {\dot M}_{BH} + {\dot M}_{rad}  = {\dot M}_{BH} + {L}_{bol} 
\eeq
${\dot M}_{BH}$ is the rate of energy absorbed by BH, and ${L}_{bol} $ is the bolometric luminosity, that is the total outflowing radiation from the BH and accretion disk system. We have Bolometric luminosity can be expressed as
\begin{eqnarray}
{L}_{bol} &=& { \cal E} \; {\dot M}_{acc} = L_{edd} \; f_{edd} \nonumber\\
{\dot M}_{BH} &=& (1- { \cal E})  \; {\dot M}_{acc}
\label{eqnaccretionterms}
\end{eqnarray}
where $L_{edd}= 10^{38}  \; \left(\frac{M}{M_\odot} \right) \; {\rm erg/s}$ is the Eddington luminosity,  $f_{edd} = L_{bol}/ L_{edd}$ is the ratio of bolometric luminosity to Eddington luminosity. $ \cal E$ is the radiative efficiency, given by
\begin{equation}
{\cal E}= 1 - \frac{{\tilde r}^{3/2} - 2 {\tilde r}^{1/2} \pm {a} }{{\tilde r}^{3/4} \left( {\tilde r}^{3/2} - 3 {\tilde r}^{1/2} \pm 2 { a} \right)^{1/2}} \; \;  \bigg|_{{\tilde r} = {\tilde r}_{ISCO}} 
\end{equation}
where $\rt = r  / GM$, and ${\cal E} ({ a}=0) \simeq 0.057$ and ${\cal E} ({ a}=1) \simeq 0.423$. Using \eqref{eqnaccretionterms} and $L_{edd}$ defined above, we end up with
\begin{equation}
{\dot M}_{BH} =\frac{ (1- { \cal E})} { { \cal E}}  \;   L_{edd} \; f_{edd} = \frac{M_{BH}} { \tau_{Acc} } \;,
\end{equation}
hence  BH growth time scale is given by
\beq
    \tau_{Acc} \simeq \frac{{\cal E}}{1-{\cal E}}  \frac{5 \cdot 10^8}{f_{edd}} \,  \; \mathrm{years}.
\eeq
For non-spinning BHs (using ${\cal E} ({ a}=0) \simeq 0.057$) we have $ \tau_{Acc} \simeq \frac{3\cdot 10^7}{f_{edd}} $ yr, and for highly spinning ones (using  ${\cal E} ({a}\sim1) \simeq 0.423$) we have $ \tau_{Acc} \simeq \frac{3.6 \cdot 10^8}{f_{edd}} $ yr.  Recall that $f_{edd}<<1$ for stellar and light BHs. In order to keep our analysis most conservative possible,  although accretion rate is usually so tiny for light BHs, ie $f_{edd}<<1$, we set $f_{edd}=1$ such that 
\begin{equation}
\tau_{Acc}=\frac{{\cal E}}{1-{\cal E}}  \; 5\cdot 10^8 \, {\rm years}
\end{equation}
For SMBHs and X-ray binaries it is possible to have $f_{edd}\sim {\cal O}(1)$ only brief time, but usually $f_{edd}<<0.01$, this gives $\tau_{Acc}>>10^{10} \, {\rm yr}$, especially for light BHs, any BH less than 10-100 solar mass.

The instability time scale, $\tau_{SR}$, is expressed as \footnote{Evolution of spin for different accretion rates in \cite{Brito:2014wla}.} 
\beq
\tau_{SR} = \frac{log (N_m)}{\Gamma_b} \, ,
\eeq
where $N_m$ is the occupation number for the corresponding state expressed as
\begin{equation}
    N_m \equiv \frac{G  M_{BH}^2 \Delta a}{m} \, ,
    \label{eqoccupationnumber}
\end{equation} 
$\Delta a$ is the spin depleted by instability, and $\Gamma_b$ is the growth rate of the bosonic field that has different mass dependencies for different spins. The field growth for scalar, vector  and tensor fields \cite{SRreview,Dias:2023ynv,Brito:2020lup} is given by
%
\beqa
&& \Gamma_{s=0} \simeq  \frac{1}{12}  \ w_H \left( r_g \, \mu \right)^8 \, \mu   \nonumber\\
&& \Gamma_{s=1} \simeq 8  \,  w_H \left( r_g \, \mu \right)^6 \, \mu \,  
\nonumber\\
&&  \Gamma_{s=2} \simeq    \, w_H \left( r_g \, \mu \right)^2 \, \mu \,   \
\label{fieldrate}
\eeqa

%


 There is a minimum spin value (and corresponding minimum mass for the boson) for the superradiance to happen for a given BH mass, given by
\begin{equation}
\Omega \geq \mu \geq \Omega_{min}= \mu_{min}\bigg|_{\tau_{Acc}=\tau_{SR}} \, 
\label{eqsrrange}
\end{equation}

\begin{figure}
\centering 
\includegraphics[width=0.495 \textwidth,angle=0]{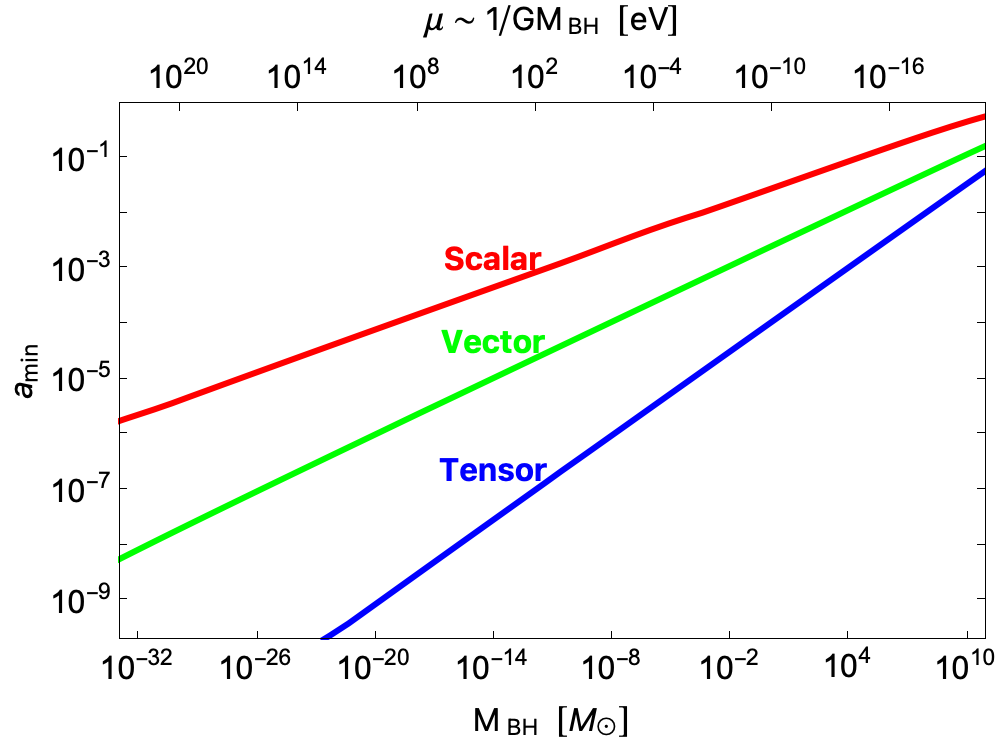}
\caption{
The minimum spin required for the black holes in the $10^{-33}-10^{11}\, M_\odot$ mass range to experience superradiance
}
\label{figminspinvsmass}
\vspace{-0.15in}
\end{figure}

 If the spin of the BH is larger than this minimum value, then it probes/constrains a range of bosonic particle masses. Hence, more rapidly rotating BHs probe a larger mass range. In Figure \ref{figminspinvsmass}, we obtain the minimum spin value for a given BH mass using \eqref{eqsrrange} by  setting $\tau_{Acc}=\frac{{\cal E}}{1-{\cal E}} \; 5 \cdot 10^8$ years. The strength of instabilities are highest for spin-2, then for vectors and weakest for scalars, however in all cases even tiny spin values are enough to experience superradiance for light black holes  (see also discussions in \cite{SRcond,SRcond1,SRreview,SRconstraints,Stott:2020gjj,Unal:2020jiy,SRnumanal}).

The field growth rate is proportional to $\propto w_H (r_g \mu)^{p_s} \mu $, where $p_s = 8, 6, 2$ for scalar, vector and spin-2. Hence with the smaller BHs we have higher rates. For example, for 1$M_\odot$ BH, horizon size is order of km, which corresponds to $10^{-5}$ seconds, and for a $10^{-15} M_\odot$ BH, it is $10^{-20}$ seconds. This parameter is compared to BH time scale which is about $10^{15}$ seconds. In the limit of minimum spin we have $\mu\sim \frac{1}{2r_g} w_{H, min}$. Hence, we have $  \tau_{Acc}   \,  / \,  r_g \, \sim \, \frac{\cal C} { w_H^p}$, where p is power of field growth set by the spin of ultralight boson ($p=10, \, 8,\, 4$ for scalar, vector and spin-2, respectively, given in eq \eqref{fieldrate}) and ${\cal C}$ is a numerical factor of order $10^{3-5}$, so we have minimum spin for superradiance for a given mass BH as 
\begin{equation}
w_{H, \, min}  \sim  \left( \frac{{\cal C} \;r_g} {  \tau_{Acc}} \right)^{1/p}
\label{eqsrcriteriaminspin}
\end{equation}
%
%

\noindent
{\bf BH Evaporation}. Black holes  their rest mass in result of quantum mechanical process, called Hawking evaporation \cite{Hawking:1974rv,Hawking:1975vcx}. 
%
%
Hawking evaporation depletes both its mass and angular momentum if radiated particles are spinning. The mass and spin evolution of the black hole is set by the temperature, spin and particle content of the vacuum at given temperature scale. 

Temperature of the black hole is given by
\begin{equation}
T_{BH} \approx  \frac{S(a)}{4\pi G M_{BH}} = 2 \,  {\rm GeV} \left( \frac{10^{-20} M_\odot}{M_{BH}} \right)
\label{eqbhtemperature}
\end{equation}
where the function, $S(a) \equiv \frac{\sqrt{1-a^2}}{1+\sqrt{1-a^2}}$, sets the temperature and evaporation rates, i.e. $S(0) =  \frac{1}{2}$ and $S(1)=0$.
\\

The degree of freedom whose mass is less than the black hole temperature, $T_{BH}$, can be produced and radiated, so BH evaporation happens faster with growing number of degrees of freedom, we will denote as $N(T_{BH})$ \cite{Page:1976df,Page:1976ki,Page:1977um}. Hence evaporation can allow us to probe heavy particles up to mass scale $T_{BH}$ \cite{MacGibbon:1990zk}, both Standard Model and  beyond Standard Model. With evaporation, mass decreases and temperature increases, therefore more degrees of freedom will join the Hawking evaporation radiation at last times. All black holes close to end of their life become smaller in mass and higher in temperature but the fraction of energy in these late moments are small fraction of total evaporation energy, so typical evaporation time is approximately set by number of degrees of freedom when BH is coldest. We assume a fixed degree of freedom for all black holes but extremely tiny ones can deviate from this time scale upto factor of ${\cal O}(0.1-10)$. Moreover, if spin becomes moderate or large, then this also modifies ${\cal O}(1)$ level the temperature and horizon size. 
\\

Time scale for evaporation can be obtained thru
\begin{eqnarray}
 {\dot M}_{BH} &=&  N(T_{BH}) \; \sigma_{SF} \; T_{BH}^4 \; A_{BH}  \nonumber\\
 &\approx& \frac{\sigma_{SF} \, N(T_{BH})}{4\pi}  \left(\frac{1}{8 \pi G M_{BH}} \right)^2  
\end{eqnarray}
where $A_{BH}=4\pi (2GM_{BH})^2 (1+\sqrt{1-a^2})$ is the area of BH 
and $\sigma_{SF}=\frac{\pi^2}{60}$ is Stefan Boltzmann constant giving
\begin{eqnarray}
\tau_{EV} &=& \frac{5120 \, \pi \, G^2 \, M_{BH}^3 }{ N(T_{BH})} \nonumber\\
& \approx & 7 \times 10^{74} \text { sec }\left(\frac{M_{BH}}{M_{\odot}}\right)^3 
\end{eqnarray}
where above we set $N(T_{BH})=10$.

Now we can compare the time scale for evaporation and superradiance as
\begin{equation}
\frac{\tau_{SR}}{\tau_{EV}}=\frac{ 1 \, \text{sec} \,  \frac{\cal C}{10^5} \, \frac{M_{BH}}{M_\odot} \, \frac{1}{w_H^p}}{ 7 \times 10^{74} \text { sec }\left(\frac{M_{BH}}{M_{\odot}}\right)^3 } \approx \frac{  10^{-33}  \, \text{sec} \,    \frac{\cal C}{10^5} \, \frac{1}{w_H^p}  \left(\frac{M_{BH}}{ \text{gr}}\right) }{ 10^{-25} \, \text{sec}  \left(\frac{M_{BH}}{ \text{gr}}\right)^3}
\end{equation}
Recall that power ``$p$'' is 10 for scalars, 8 for vectors and 4 for spin-2,
hence we find that as long as BH is rotating rapidly the superradiance
occurs much faster than evaporation at any BH mass BH. Hence, superradiance
drives the spin of the black hole to $w_{H, min}$ given
in \eqref{eqsrcriteriaminspin} or in Fig.~\ref{figminspinvsmass}, and when the spin decreases, the superradiance rate strongly gets suppressed. 

The phase boundary between the superradiance and evaporation is

\begin{equation}
 w_{H, min} \approx \left(  10^{-8} \, \frac{\cal C}{10^5} \, \left(\frac{M_{BH}}{ \text{gr}}\right)^{-2} \right)^{1/p} 
 \label{eqsrevapboundary}
\end{equation}
\\

 \begin{figure}
\centering 
\includegraphics[width=0.495\textwidth,angle=0]{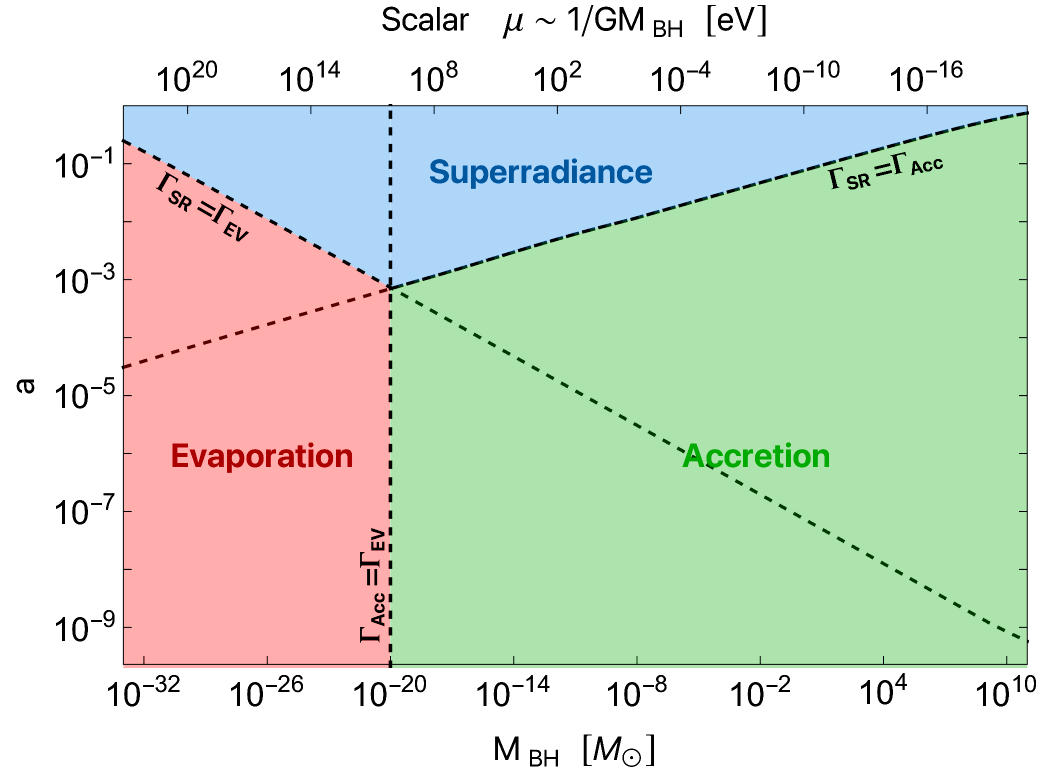}
\vspace{0.25in}

\includegraphics[width=0.495\textwidth,angle=0]{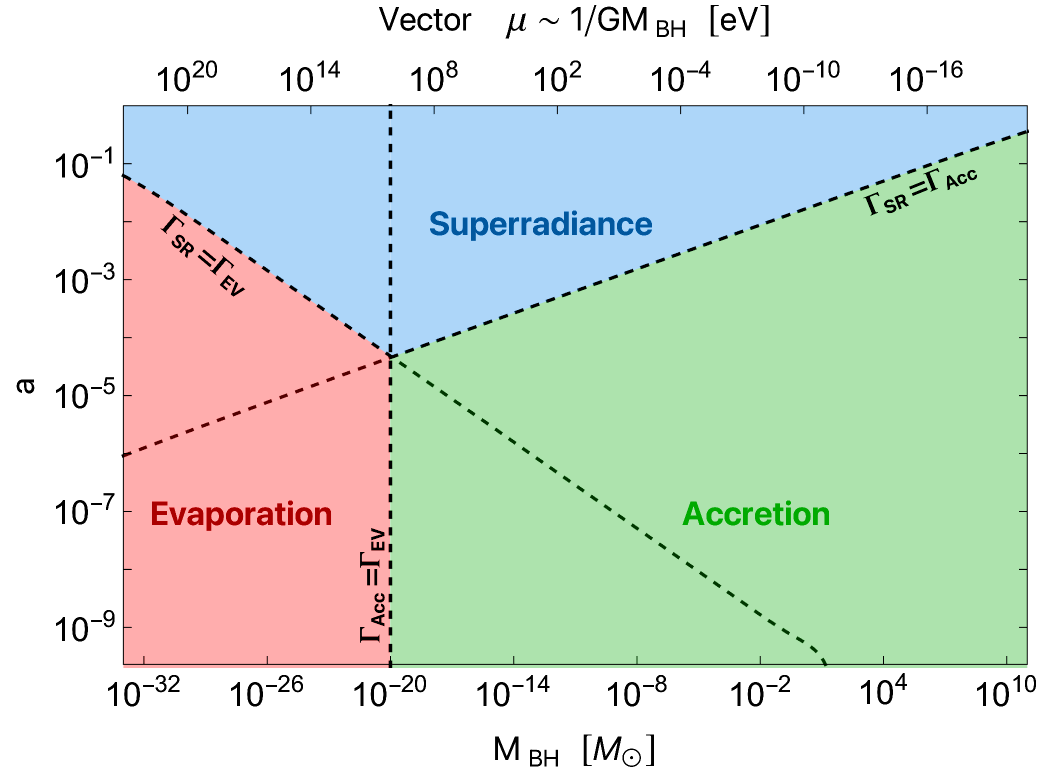}
\vspace{0.25in}

\includegraphics[width=0.495\textwidth,angle=0]{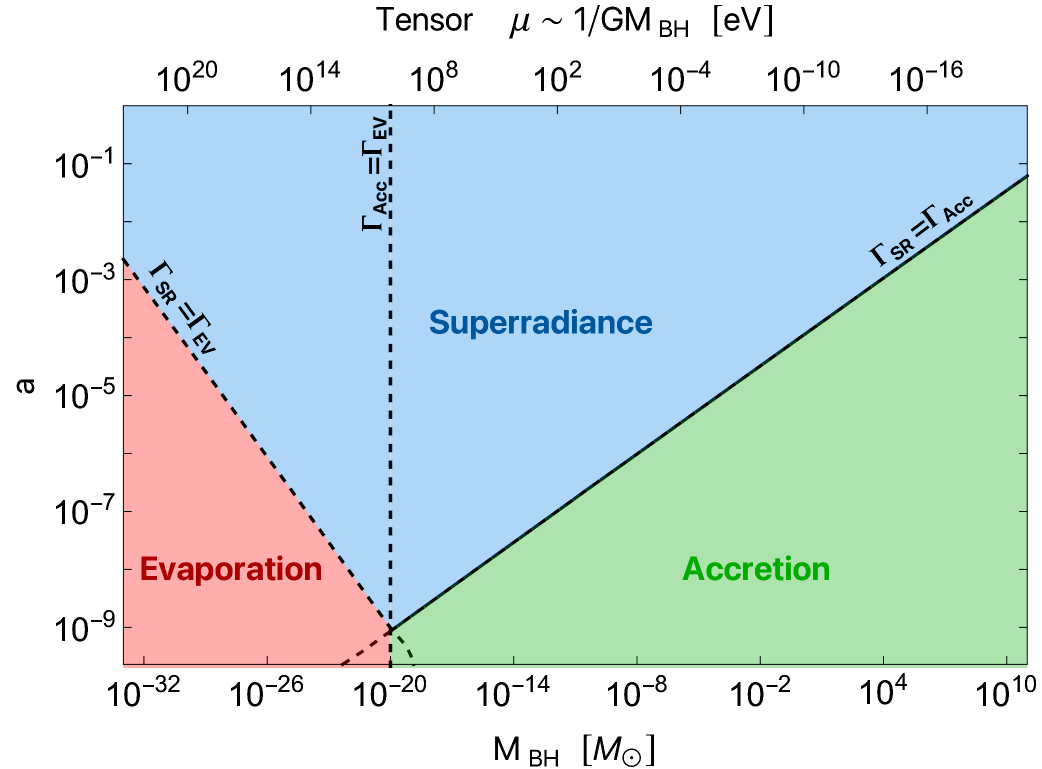}
\vspace{0.1in}
\caption{
Phase Diagram of Black Holes for (Eddington) accretion, evaporation, (scalar, vector and tensor) superradiance
}
\label{figphasediagram}
\vspace{-0.25in}
\end{figure}

\noindent
{\bf Phase Boundary of Evaporation and Acretion}.
We give more complete phase boundary between evaporation and accretion ($ \tau_{EV} \sim \tau_{Acc} $), including effective degree of freedom and spin, as 
\begin{equation}
\tau_{EV} = \frac{4(1-a^2)^2}{(1+\sqrt{1-a^2})^2} \frac{ 10^{-25}sec (M_{BH}/gr)^3 }{N(T_{BH})/10} \simeq  \tau_{Acc} 
\label{eqaccevapboundary}
\end{equation}
For moderate and small spin values, we have only mass dependent relation 
\begin{equation}
M_{BH} \approx 2\cdot 10^{13} gr \approx 10^{-20} M_\odot
\label{eqaccevapboundary2}
\end{equation}
For lower mass black holes, more degrees of freedom contributes to radiation, $N(T_{BH})$, and this shortens the evaporation time, and dof change is neglected in Fig. \ref{figphasediagram}. 


\vspace{0.2cm}
\noindent
{\bf Phase Diagram of Black Holes}. We obtain the phase diagram of black hole evolution via distinct processes that are accretion, Hawking radiation, and superradiance. When the time scale for any of those processes are much shorter than the others, that process governs the BH evolution. 
 If the mass of the black hole is less than $10^{-20} M_\odot$, for large spins, also if there exists corresponding bosonic particle, black holes superradiate and decrease their spin to small values which is given in equation \eqref{eqsrevapboundary}, and then evaporation dominates the evolution. If the mass of the black hole is more than $10^{-20} M_\odot$, then  for large spins, also if there exists corresponding bosonic particle, black holes superradiate and decrease their spin to small values which is given in equation \eqref{eqsrcriteriaminspin}, and then accretion dominates the evolution if it is order of Eddington rate. If accretion rate is smaller than Eddington rate, then black hole keeps superradiating until accretion and supperradiation balances each other.

%


 \begin{figure*}
\centering 
\includegraphics[width=0.99 \textwidth,angle=0]{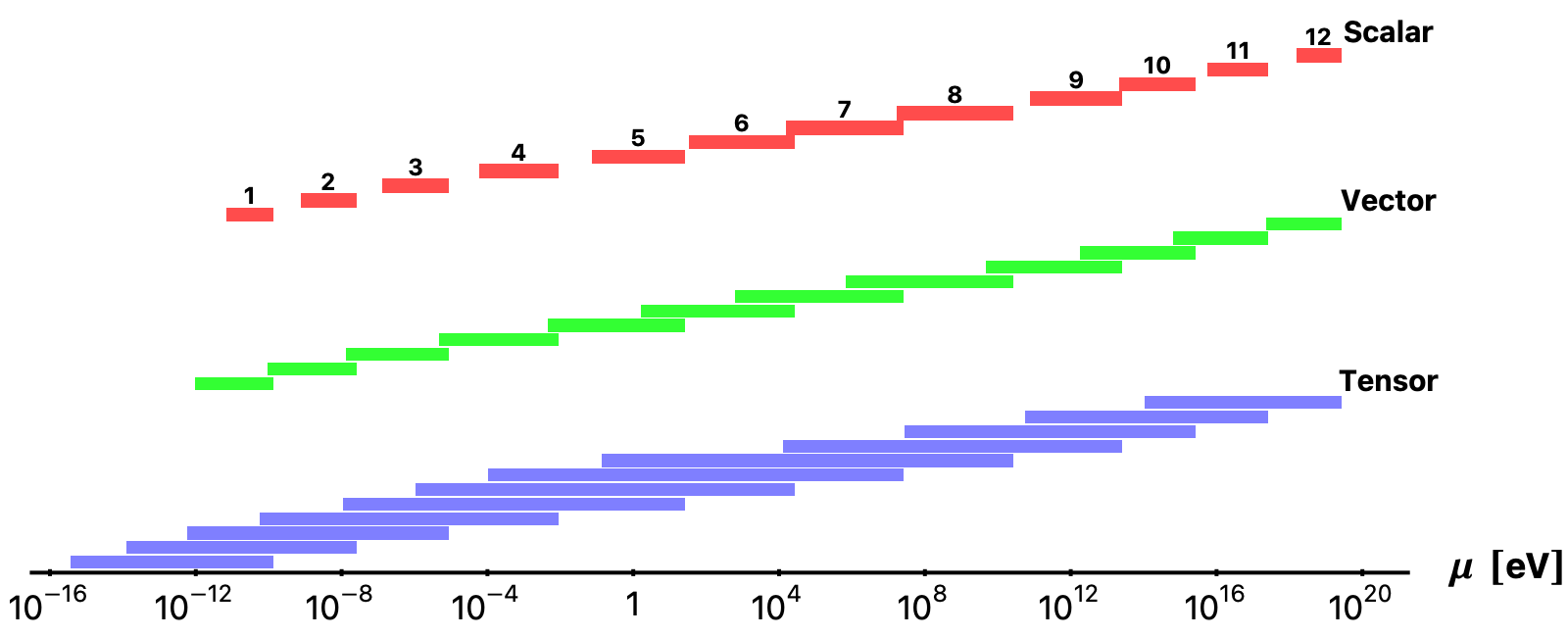}
\caption{
Scalar, Vector and Spin-2 (Tensor) particles probed by superradiance of light (less than solar mass) BHs
}
\label{figall}
\vspace{-0.15in}
\end{figure*}

We show in Figure \ref{figphasediagram}, the generic phase diagram for black holes in the mass-spin plane, for the mass range of black holes $10^{-33}-10^{11} M_\odot$. Top panel is for scalar supperradiance, middle for vector supperradiance and bottom for spin-2 supperradiance.  There are 3 dashed lines indicating similar rates for I) $\Gamma_{SR} = \Gamma_{EV}$ given in \eqref{eqsrevapboundary}, II) $\Gamma_{SR} = \Gamma_{Acc}$ given in \eqref{eqsrcriteriaminspin}, and III) $\Gamma_{Acc} = \Gamma_{EV}$) given in \eqref{eqaccevapboundary}, 3 colorful regions (superradiance dominated shown in blue, ie. superradiance is faster than other two processes; evaporation dominated shown in red, ie. evaporation is faster than other two processes; accretion dominated shown in green, ie. accretion is faster than other two processes). Observe that colorful regions (blue, red and green) are also divided into 2 pieces in which subdominant processes have different rates. For superradiance (blue), right sub-region corresponds to $\Gamma_{SR} > \Gamma_{Acc} > \Gamma_{EV}$, and left sub-region corresponds to $\Gamma_{SR} > \Gamma_{EV} > \Gamma_{Acc} $. Similarly for the evaporation (red), upper sub-region corresponds to $\Gamma_{EV} > \Gamma_{SR} > \Gamma_{Acc} $, and lower sub-region corresponds to $\Gamma_{EV} > \Gamma_{Acc} >\Gamma_{SR}$. Finally for the accretion (green), upper sub-region corresponds to $\Gamma_{Acc}  > \Gamma_{SR}  > \Gamma_{EV} $, and lower sub-region corresponds to  $\Gamma_{Acc}  > \Gamma_{EV} > \Gamma_{SR}  $.

\vspace{0.2cm}

\noindent
{\bf Light Black Holes as Boson Probes.} 
Black holes can interact with scalar (or pseudoscalar such as axionlike), vector (dark photon and/or photon with effective mass via interactions with plasma) and spin-2 particles, and the interaction strength is set by the instability rate. For larger instability rates (especially for light BHs) even small spin values are enough for BH to enter superradiance. This minimum spin value also defines a minimum boson mass the BH can interact, hence for high instability rate smaller mass bosons could interact with BH. The upper mass limit for the boson is set by the maximum allowed spin value which is $a=1$ corresponding to speed of light. Therefore, BH probes a range of boson mass starting from the minimum boson mass corresponding to  minimum superradiance spin until maximum boson mass corresponding to the nearly speed of light rotation. If the spin value is large, this allows probing large range of boson masses. 

For a given BH mass and spin, there is a corresponding mass range that can be probed for scalar, vector and tensor (spin-2) particles. The superradiance bounds for black holes heavier than solar mass (stellar to supermassive) have been studied in detail \cite{SRcond,SRcond1,SRreview,Dias:2023ynv,Stott:2020gjj,Unal:2020jiy,Roy:2021uye,Chen:2022nbb,Chen:2021lvo}, hence we focus on the parameter space probed by light primordial black holes (subsolar mass) that can have a rich phenomenology including dark matter, dark radiation, induced gravitational wave, superradiance and baryogengesis scenarios \cite{Khlopov:1985fch,Rosa:2017ury,Bernal:2021yyb,Branco:2023frw,Richarte:2021fbi,Mazde:2022sdx,Calza:2023rjt,Hooper:2019gtx,Papanikolaou:2020qtd,Bernal:2022oha,March-Russell:2022zll,Gehrman:2022imk}.  We assume hypothetical detection of spinning ($a \sim 0.7$)  black holes, ${\rm BH_{1-12}}$, with masses
\begin{eqnarray}
M_{BH}/M_\odot= \{0.2, \,10^{-3}, 10^{-5.5},  \,10^{-8.5}, \, 10^{-12}, \,10^{-15},  \nonumber\\
 \,10^{-18}, \,10^{-21}, \,10^{-24}, \,10^{-26}, \,10^{-28}, \,10^{-30}\} \,   \nonumber  
\end{eqnarray}
%
We show in Figure \ref{figall} that each spinning BH will allow us to probe a mass range of nearly 2-9 decades (an example model for wide range of PBHs is given in Ref. \cite{Franciolini:2022pav} and late formation in \cite{Chakraborty:2022mwu}.

Another way to probe heavy Standard Model and beyond Standard Model particles is the evaporation of light primordial black holes. The rate of black hole evaporation depends on all the massless and massive degrees of freedom, $N(T_{BH})$  given in \eqref{eqbhtemperature}. Hence light primordial black holes can probe heavy dark sectors, sterile neutrinos, heavy neutral particles and supersymmetric particles via evaporation (both fermons and bosons)  and via superradiance (bosons).
\vspace{0.2cm}

\noindent{\bf Bounds on Self-Interactions of Bosonic Fields.}  We start with the sinusoidal axion potential, neglecting higher harmonics, $V=\Lambda^4 (1- \cos \frac{\phi}{f_a} )$, where $\phi$ is the axion, and $f_a$ is the axion decay constant. The mass is given by $\mu= \Lambda^2/f_a$, and self-interaction by $\lambda= \frac{\Lambda^4}{f_a^4}$. In the case of strong external and self-interaction, superradiance can be prevented \cite{nosuperradiancebyinteractions1,nosuperradiancebyinteractions2,Cannizzaro:2022xyw}.

In the presence of self-interaction, superradiance condition can be expressed as \cite{Yoshino:2012kn,SRcond,SRcond1,SRreview}   (see also \cite{Babichev:2016bxi,spin2SR,Poddar:2019zoe})
\begin{equation}
\Gamma_{SR} \; \tau_{Acc} \, \left(N_{BOSE}/N_{m} \right) > log N_{BOSE} \, ,
\label{eqnovasuperradiance}
\end{equation}
where $N_{m}$ being the occupation number
\begin{equation}
    N_{BOSE} \simeq 5 \cdot 10^{44} \frac{n^4}{(r_g \, \mu)^3} \left ( \frac{M_{BH}}{10^{-8} \,M_\odot} \right)^2 \left(\frac{f_a}{ 10^{10} \, {\rm GeV}}   \right)^2
\end{equation}
where the prefactor is obtained via numerical analysis, $\sim5$, in Ref. \cite{Yoshino:2012kn}. 

\begin{figure}
\centering 
\includegraphics[width=0.489\textwidth,angle=0]{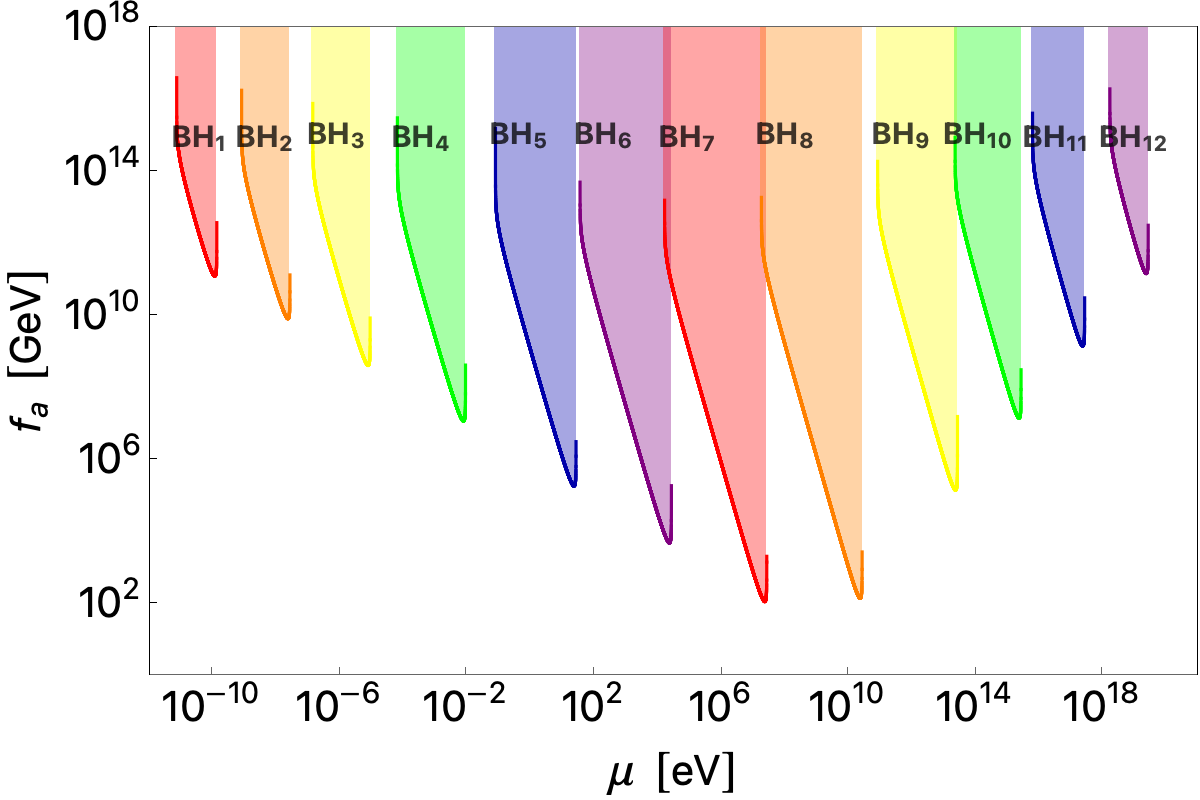}
\includegraphics[width=0.5\textwidth,angle=0]{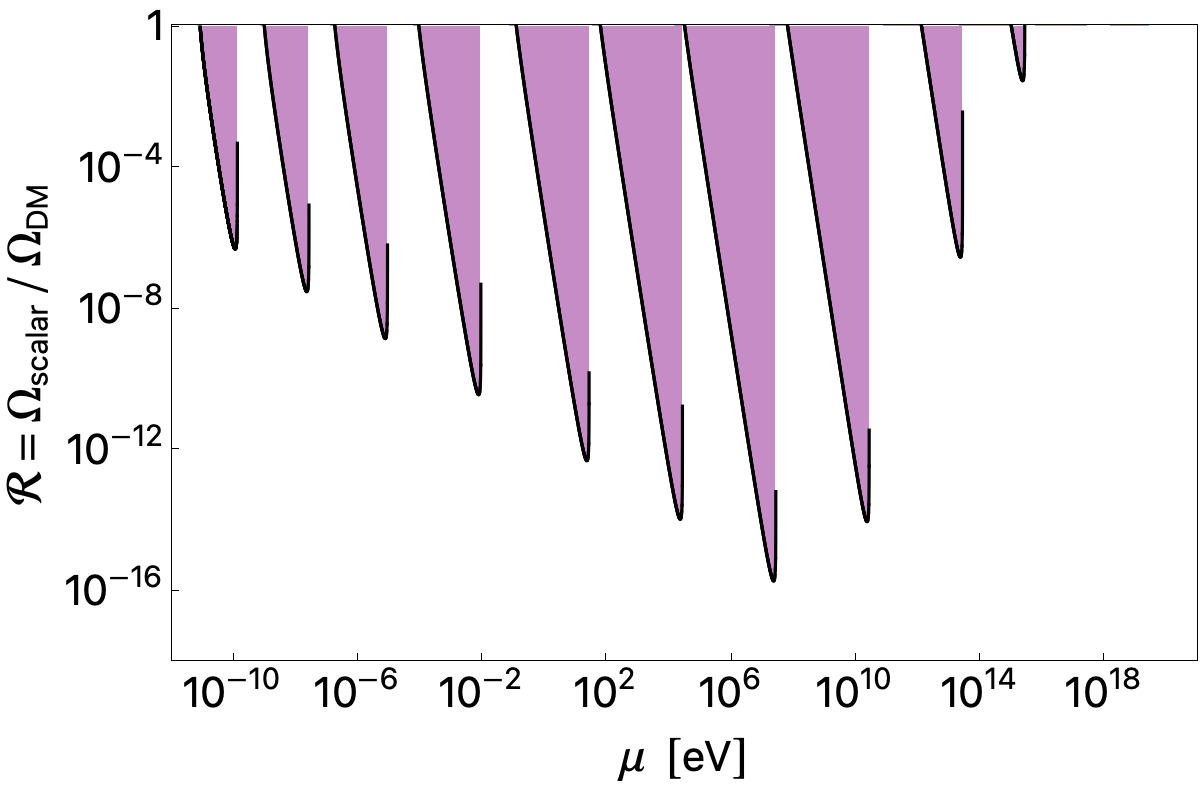}
\caption{
Top: Bounds on ${\rm f_{axion}}$ in the presence of spinning light BHs as a function of scalar/axion mass; \; Bottom:  Bounds on the energy density as a function of scalar/axion mass
}
\label{figselfinteractionenergydensityscalar}
\vspace{-0.15in}
\end{figure}

\begin{figure}
\centering 
\includegraphics[width=0.489\textwidth,angle=0]{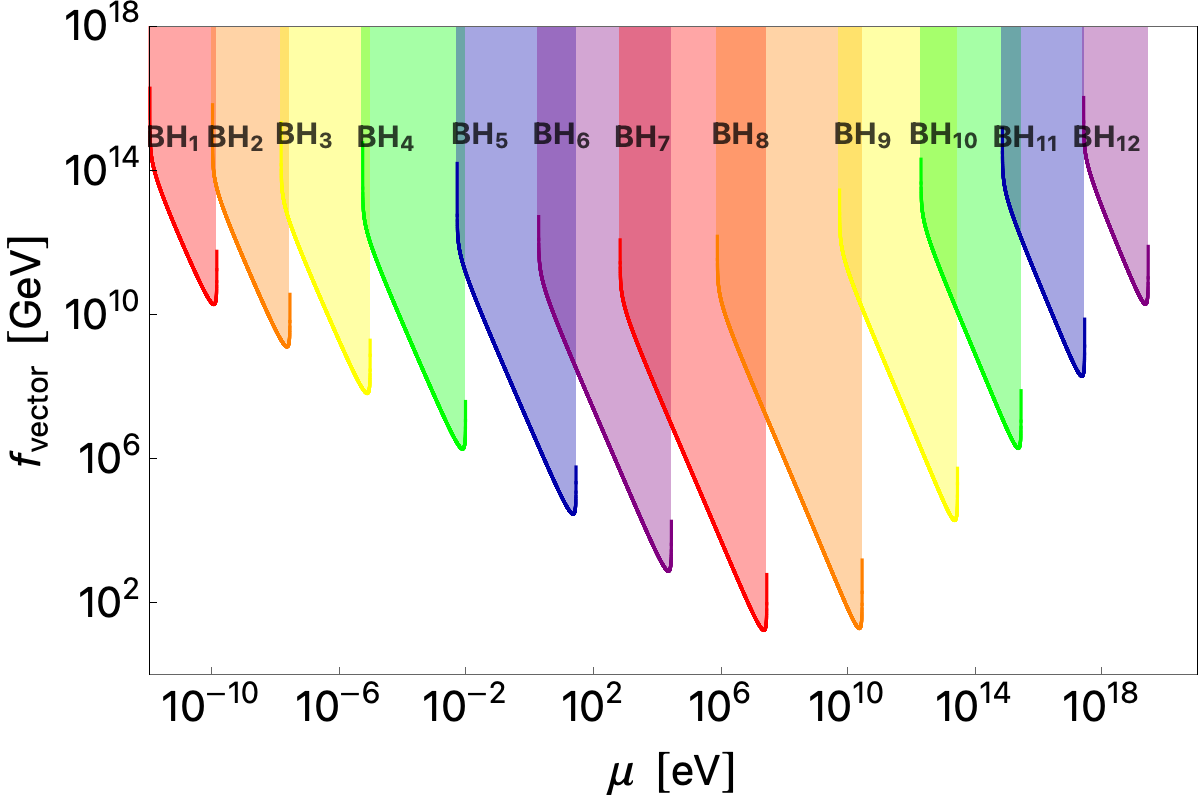}
\includegraphics[width=0.5\textwidth,angle=0]{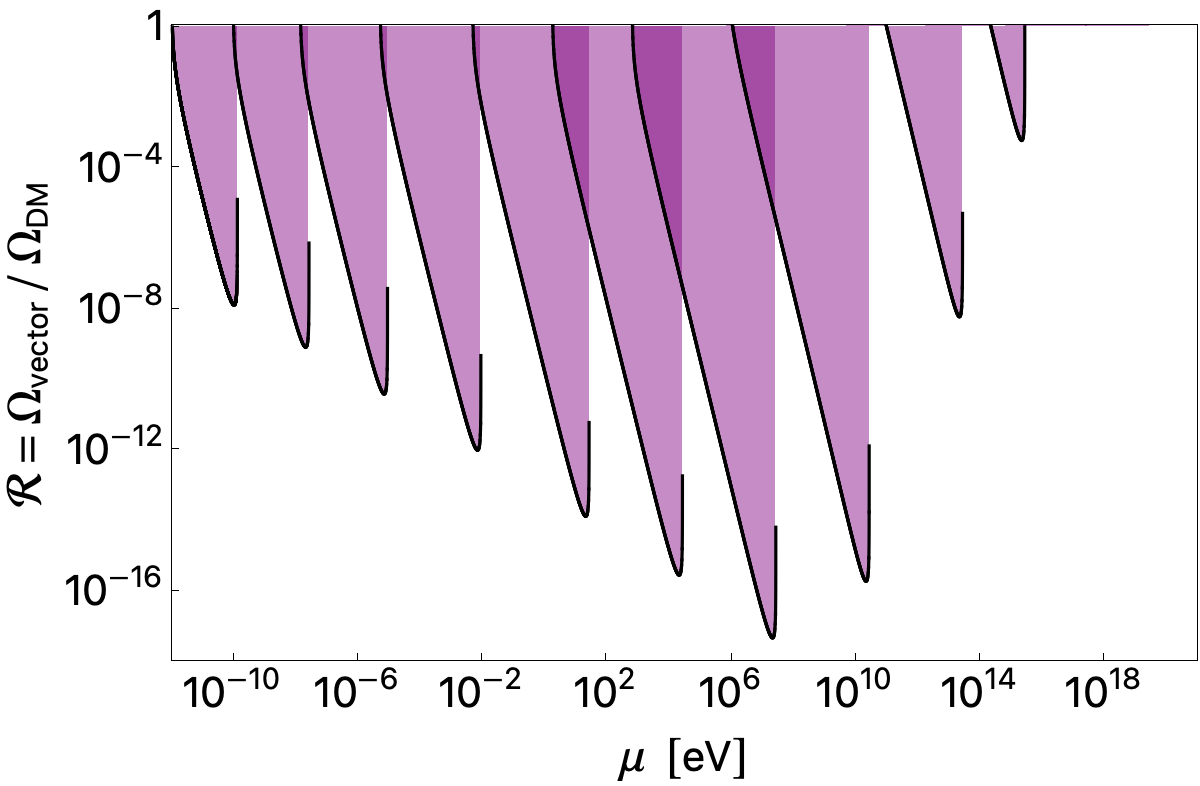}
\caption{
Top: Bounds on ${\rm f_{vector}}$ in the presence of spinning light BHs as a function of vector mass; \; Bottom:  Bounds on the energy density as a function of vector mass
}
\label{figselfinteractionenergydensityvector}
\vspace{-0.15in}
\end{figure}

\begin{figure}
\centering 
\includegraphics[width=0.489\textwidth,angle=0]{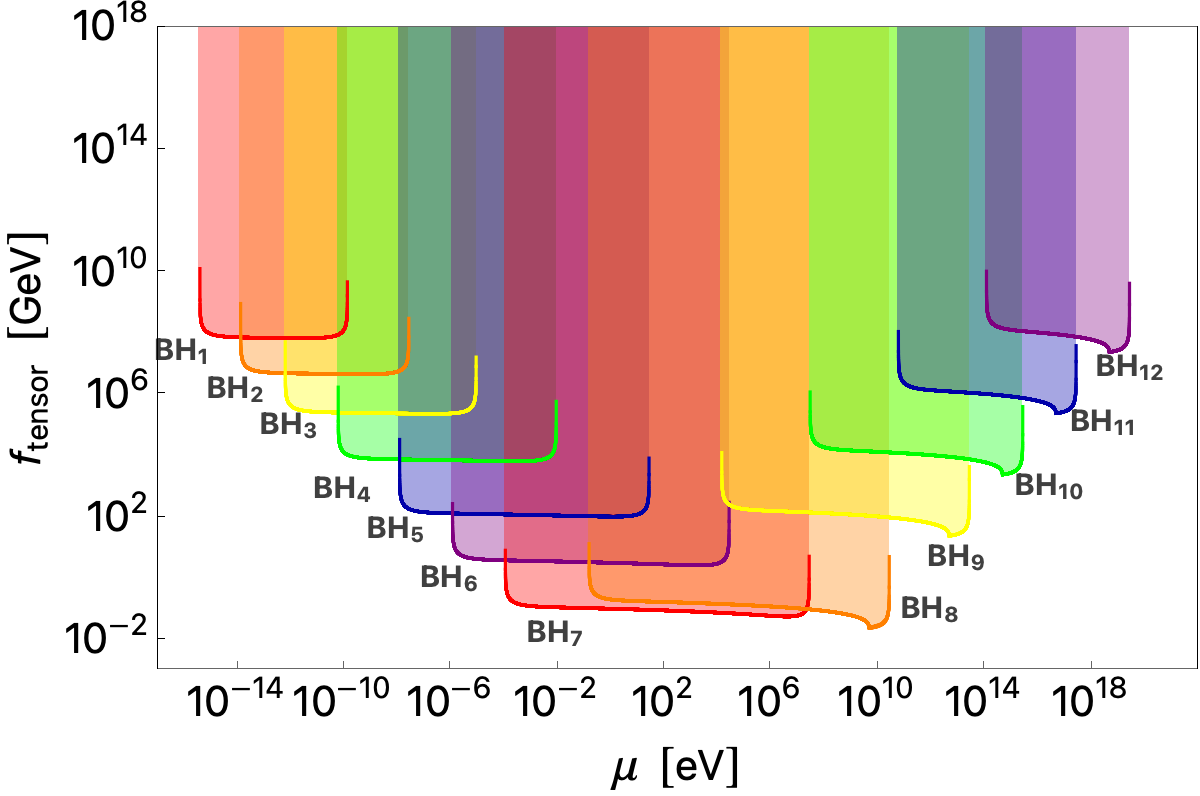}
\includegraphics[width=0.5\textwidth,angle=0]{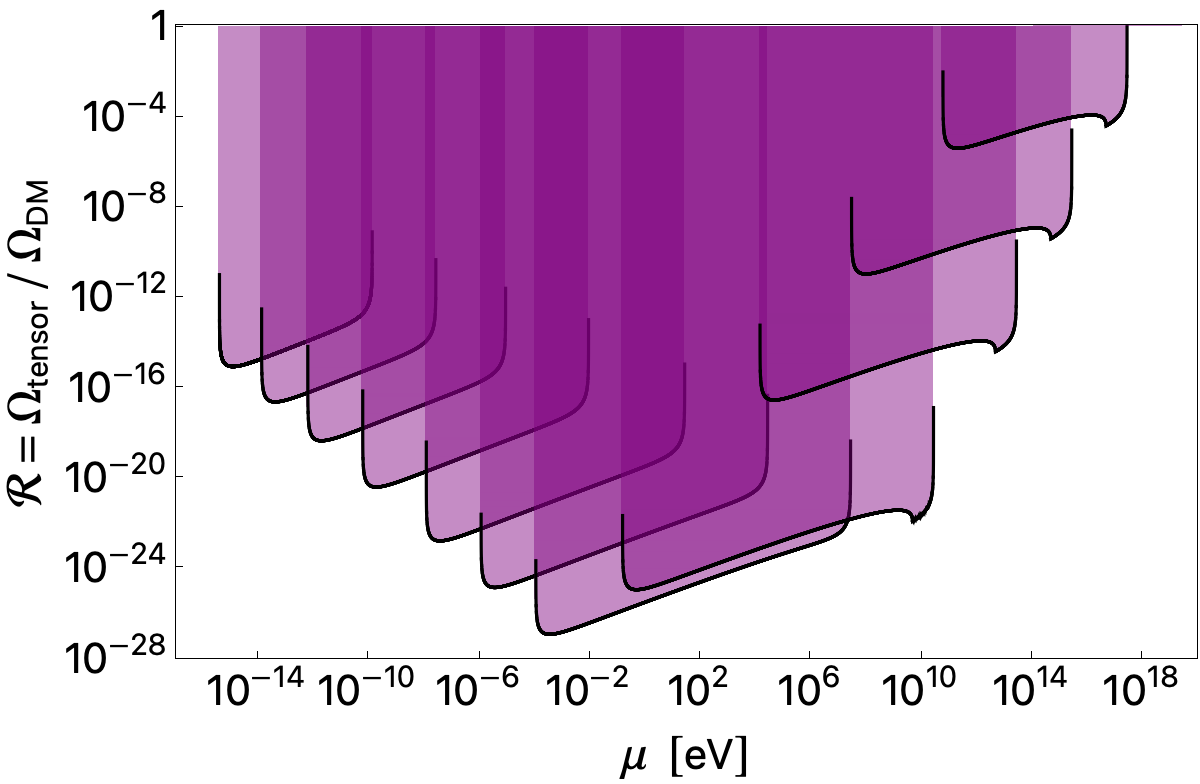}
\caption{
Top: Bounds on ${\rm f_{tensor}}$ in the presence of spinning light BHs as a function of spin-2 mass; \; Bottom:  Bounds on the energy density as a function of spin-2 mass
}
\label{figselfinteractionenergydensitytensor}
\vspace{-0.15in}
\end{figure}

If BH is spinning fast and superradiance is not observed, then one option is that there is no such particle in the corresponding mass range. The other option for the non-observation of the superradiance could be self (and/or external) interactions. The growing self-interactions with decreasing decay constant $f_a$, could prevent the BH from superradiance. In such a case, one can derive limits on the decay constant of the axion (or vector or spin-2) as in top panel of Figure \ref{figselfinteractionenergydensityscalar} (\ref{figselfinteractionenergydensityvector} or \ref{figselfinteractionenergydensitytensor}). A more in depth study is performed in \cite{Baryakhtar:2020gao} that the bounds on the self-interaction can be derived via interactions of different energy levels.
%
%
%
%
%

\vspace{0.2cm}
\noindent{\bf Standard Model and Beyond Bosons \& Higgs Self-Coupling.}
As the mass of the BH decreases, its horizon size decreases and the mass of the probed boson masses increase. Lightest massive particles in the Standard Model (SM) are neutrinos and leptons, which are fermions. Lightest bosons of Standard Model are mesons and massive electroweak bosons. Black holes lighter than $10^{15} \, {\rm grams}= 10^{-18} M_\odot$ can probe $10^{8} \, {\rm eV}$, pions \cite{Ferraz:2020zgi} and heavier bosons ($K, D, B, \eta, \, \eta', \, \eta_c, \eta_b$) \cite{ParticleDataGroup:2022pth}. If there exist black holes lighter than $10^{11} {\rm grams}=10^{-22}M_\odot$, they can probe $10^{12} \, {\rm eV}$ and heavier bosons, including electroweak vector bosons and Higgs field. 
Note that in SM only electrons, protons, neutrinos and photons  can stay stable for long periods, hence those particles decay in their dominant channels, typically producing gamma-ray photons and neutrinos.
Producing such unstable SM and beyond particles requires the superradiance rates  to be faster than decay rate ($\Gamma_{SR} > \Gamma_{decay}$), where balance law is given by 
\begin{equation}
N \propto \; {\rm exp} \big[ t \left(  \Gamma_{SR} - \Gamma_{decay} \right)  \big]
\end{equation}
In colliders, the Higgs particle lifetime is measured as 
$1/ \Gamma(H)\sim 1.6 \cdot 10^{-22} \, {\rm sec}$. 
The lifetime (inverse decay rates) of the mesons \cite{ParticleDataGroup:2022pth} vary, typically in the range $10^{-25}-10^{-8} \; {\rm sec}$. For pion and meson production, i.e. the superradiance rate to be faster than the decay rate, black holes spinning faster than about 0.15 needed; and for the Higgs production, black holes spinning larger than about 0.7 needed. However, Higgs self-interactions can also prevent superradiance. Even if such particles are produced, when superradiance slows down or stops, these particles decay via their dominant channels and produce light stable particles such as neutrinos and high energy photons.

\vspace{0.2cm}
\noindent{\bf The Fraction of Dark Matter in Scalar, Vector and Tensor Fields.}
%
%
%
%
Dark matter can have scalar, vector and spin-2 components \cite{Hu:2000ke,Hui:2016ltb,Marsh:2015xka,Marzola:2017lbt}. In order to derive its current energy density we focus on two times: 
\\
i) Start of rolling time: $H\sim \mu$. At early times, when the Hubble parameter is larger than its mass, scalar field does not roll, so rolling time is
\begin{equation}
    \rho=3H_{roll}^2 M_p^2 \sim T_{roll}^4 \Rightarrow H_{roll}\sim \mu\sim T_{roll}^2/M_p
\end{equation}
ii) Matter-radiation equality : When $\mu>H$, field oscillates and behaves like non-relativistic dust. Around matter-radiation equality, S being scale factor, the fraction of dark matter in terms of axion is given as
%
\begin{eqnarray}
{\cal R} \equiv \frac{\Omega_{scalar}}{\Omega_{DM}} &=& \frac{\rho_{axion}}{\rho_{radiation}} \bigg|_{t_{eq}} \sim \frac{\Lambda^4}{T_{roll}^4} \frac{ S_{eq}}{S_{roll}} \sim \frac{ \mu^2 f_a^2}{T_{roll}^3 T_{eq}}
\nonumber\\ &&\Rightarrow \mu^2 f_a^2 \sim  {\cal R} \;  ( \mu \, M_p)^{3/2} \, T_{eq}
\end{eqnarray}
They result in
\begin{equation}
      {\cal R} \equiv \frac{\Omega_{scalar}}{\Omega_{DM}}  \sim \, \left( \frac{\mu}{10^{-1} \, \mathrm{eV}}\right)^{1/2} \left( \frac{f_a}{10^{12} \, \mathrm{GeV}} \right)^2 \,.
       \label{eqendensity}
\end{equation}

If spinning BHs are detected, then using the upper bounds on the decay constant, we can derive upper limits on the energy density as a function of particle mass via \eqref{eqendensity}. In bottom panel of Figure \ref{figselfinteractionenergydensityscalar} (also Figure \ref{figselfinteractionenergydensityvector} and \ref{figselfinteractionenergydensitytensor} ), we show the bounds on the ratio of energy density in scalars (vectors and spin-2) to total dark matter energy density as a function of particle mass. We note that our analysis is conservative since focuses only for the most unstable superradiance state, and it is possible to probe higher mass particles (with higher azimuthal number) but with less constraint, for a given BH mass and spin. In our conservative analysis, we find that if highly spinning light BHs exist, and quenching of superradiance is due to self-interactions, then scalar particles in the corresponding boson mass range can constitute at most $\Omega_{scalar} \, / \, \Omega_{DM}< 10^{-16}-10^{-3}$ of dark matter in the mass range $10^{-10}-10^{15}$eV; vectors can onstitute at most $\Omega_{vector} \, / \, \Omega_{DM}< 10^{-18}-10^{-3}$ of dark matter in the mass range $10^{-10}-10^{15}$eV; and spin-2  can constitute at most $\Omega_{spin-2} \, / \, \Omega_{DM}< 10^{-27}-10^{-3}$ of dark matter in the mass range $10^{-10}-10^{15}$eV.

\vspace{0.2cm}
\noindent
{\bf Conclusions.} We study the evolution of (primordial) black holes under accretion, evaporation and superradiance, and obtain the generic phase diagram under these processes for the mass range from $10^{-33}M_\odot\sim 1$ gram to $10^{11}M_\odot\sim  10^{44}$ grams.  We derive the phase boundaries between  the superradiance (for scalar, vector and tensor perturbations), accretion and evaporation. We explore the implications of our results on primordial black holes, and bosons and fermions of Standard Model and beyond SM particles  (mass parameter space, the self-interaction strength and energy density). We summarize our results as follows :
\\

\noindent 1)  We obtain the phase diagram of black holes in the black hole mass range, $(10^{-33}-10^{11}) M_\odot$, for processes accretion, evaporation (for both fermions and bosons) and superradiance (for bosons) corresponding to particles in the mass range $(10^{-21}-10^{21})$ eV. The rate for accretion is assumed to be approximately Eddington rate in this work and only radiation efficiency dependent which is function of time. Superradiance rate depends on the mass of the black hole and type of particle interacting with black hole, namely scalar, vector or spin-2. It is highest for spin-2, then for vector and slowest for scalars. On the other hand, evaporation rate is set by the temperature of the black hole and horizon size which allows both bosonic and fermonic degrees of freedom to contribute Hawking radiation, whose mass   is less than or order of the horizon temperature.  For BHs heavier than $10^{-20}M_\odot$, superradiance and accretion processes are more relevant, for BHs lighter than $10^{-20}M_\odot$ all three processes are relevant and we find three phase boundaries for these 3 phases, the boundary for superradiance-accretion given in eq. \eqref{eqsrcriteriaminspin}, for superradiance-evaporation given in eq. \eqref{eqsrevapboundary}, and for accretion-evaporation given in eq. \eqref{eqaccevapboundary}.
\\

\noindent  2) Black holes heavier than solar mass can probe at most $10^{-12}$ eV bosons. Light black holes can probe heavier phenomenologically interesting bosons via superradiance and evaporation, and fermions via evaporation,  such as QCD axion, Higgs, dark photon, spin-2 dark matter, sterile neutrino, supersymmetric particles etc. Superradiance instability is stronger for light black holes $(10^{-33}-1) M_\odot$. We find that superradiance can start from tiny spin values as low as $10^{-4}-10^{-2}$ for scalars, $10^{-6}-10^{-4}$ for vectors and $10^{-9}-10^{-4}$ for spin-2 (Figure \ref{figminspinvsmass}). Since light BHs are unstable for such perturbations and sensitive probes of bosons, a single BH can probe 2-9 decades of mass range for corresponding bosons. Evaporation gets stronger with decreasing mass (and low spin values) as the horizon temperature increases, and it is especially relevant for black holes  lighter than $10^{-20}M_\odot$, which correspond to particles heavier than GeV  (Figure \ref{figphasediagram}).
\\

\noindent 
3) Standard Model and beyond can be probed by evaporation ( both fermions and bosons) and superradiance (bosons).  With  $10^{-22}M_\odot $ and lighter black holes, we can probe the nature of the Higgs field, i.e. mass and self- and external interactions. Similarly with $10^{-18}M_\odot $ and lighter black holes, pions and other mesons can be  probed.  To produce Higgs, pions, mesons and other heavy particles with high decay rates, black holes need to spin moderately or rather fast. These heavy particles are unstable (SM ones), hence they decay to stable light particles after superradiance slows down. 
\\

\noindent 
4) Due to superradiance instability, evaporation and negligible accretion, independent from their formation mechanism and initial spin value, light and ultralight primordial black holes rapidly loose their spin. After they reach to spin values at the superradiance and evaporation phase boundary, they continue their evolution with Hawking radiation which further depletes their mass and spin. Hence either due to beyond Standard Model bosons and/or effectively massive photons in plasma and/or rapid evaporation, we expect light and ultralight primordial black holes to be nearly non-spinning ( Figure \ref{figphasediagram}).
\\

\noindent 
5) Detection of superradiance can imply existence of such particles, but if there are spinning light BHs implying the non-existence of superradiance, this can also lead to two conclusions: i) The scalar, vector, tensor particles do not exist in that relevant mass range and ruled out (see Figure \ref{figall}), ii) Such particles have strong self-interactions which prevent them from experiencing superradiance. In such a case, strong self-interactions requires upper bounds on axion (or vector or spin-2) decay constant and the energy density (Figures \ref{figselfinteractionenergydensityscalar}, \ref{figselfinteractionenergydensityvector}, \ref{figselfinteractionenergydensitytensor}), and the bosons in the corresponding mass range can not contribute more than $10^{-18}-10^{-3}$ fraction of the dark matter in the boson mass range $10^{-10}-10^{15}$ eV.

\vspace{0.25cm}
\noindent
{\bf  Acknowledgements.} I am grateful to Richard Brito, Vitor Cardoso, Gabriele Franciolini, Ely D. Kovetz, Sarah Libanore,  Paolo Pani, Debanjan Sarkar for the discussions and suggestions on the manuscript, also Theodoros Papanikolaou and João Rosa for the comments on evaporating BHs and the superradiance of Standard Model particles, Andrea Caputo for the scattering of low-density particles, the referee for his/her detailed feedback and suggestions for evaporation effects, and my family for their support.  I dedicate this work to Gökçe Erhan, and Cihan, G\"ozde and {\c S}imal G\"ulbudak. I would like to thank INFN Padova, 
UAM IFT, 
Technion,  
Dieter Schwarz Stiftung, Lindau Nobel Laureates Meeting, Evelin \& Erwin Brugger and Efrat Forti for their hospitality and support during this work. This research is supported by the Kreitman fellowship of the BGU, and the Excellence fellowship of the Israel Academy of Sciences and Humanities, and the Council for Higher Education.








\begin{thebibliography}{99}



\bibitem{Carr:2020gox}
B.~Carr, K.~Kohri, Y.~Sendouda and J.~Yokoyama,
``Constraints on primordial black holes,''
Rept. Prog. Phys. \textbf{84}, no.11, 116902 (2021)
doi:10.1088/1361-6633/ac1e31
[arXiv:2002.12778 [astro-ph.CO]].


\bibitem{Carr:2020xqk}
B.~Carr and F.~Kuhnel,
``Primordial Black Holes as Dark Matter: Recent Developments,''
Ann. Rev. Nucl. Part. Sci. \textbf{70}, 355-394 (2020)
doi:10.1146/annurev-nucl-050520-125911
[arXiv:2006.02838 [astro-ph.CO]].

\bibitem{Green:2020jor}
A.~M.~Green and B.~J.~Kavanagh,
``Primordial Black Holes as a dark matter candidate,''
J. Phys. G \textbf{48}, no.4, 043001 (2021)
doi:10.1088/1361-6471/abc534
[arXiv:2007.10722 [astro-ph.CO]].

\bibitem{Sasaki:2016jop}
M.~Sasaki, T.~Suyama, T.~Tanaka and S.~Yokoyama,
Phys. Rev. Lett. \textbf{117}, no.6, 061101 (2016)
[erratum: Phys. Rev. Lett. \textbf{121}, no.5, 059901 (2018)]
doi:10.1103/PhysRevLett.117.061101
[arXiv:1603.08338 [astro-ph.CO]].

\bibitem{Raidal:2017mfl}
M.~Raidal, V.~Vaskonen and H.~Veerm\"ae,
JCAP \textbf{09}, 037 (2017)
doi:10.1088/1475-7516/2017/09/037
[arXiv:1707.01480 [astro-ph.CO]].

\bibitem{Ali-Haimoud:2017rtz}
Y.~Ali-Ha\"\i{}moud, E.~D.~Kovetz and M.~Kamionkowski,
Phys. Rev. D \textbf{96}, no.12, 123523 (2017)
doi:10.1103/PhysRevD.96.123523
[arXiv:1709.06576 [astro-ph.CO]].

\bibitem{Aggarwal:2020olq}
N.~Aggarwal, O.~D.~Aguiar, A.~Bauswein, G.~Cella, S.~Clesse, A.~M.~Cruise, V.~Domcke, D.~G.~Figueroa, A.~Geraci and M.~Goryachev, \textit{et al.}
Living Rev. Rel. \textbf{24}, no.1, 4 (2021)
doi:10.1007/s41114-021-00032-5
[arXiv:2011.12414 [gr-qc]].

\bibitem{Franciolini:2022htd}
G.~Franciolini, A.~Maharana and F.~Muia,
``Hunt for light primordial black hole dark matter with ultrahigh-frequency gravitational waves,''
Phys. Rev. D \textbf{106}, no.10, 103520 (2022)
doi:10.1103/PhysRevD.106.103520
[arXiv:2205.02153 [astro-ph.CO]].



\bibitem{Saito:2008jc}
R.~Saito and J.~Yokoyama,
``Gravitational wave background as a probe of the primordial black hole abundance,''
Phys. Rev. Lett. \textbf{102}, 161101 (2009)
[erratum: Phys. Rev. Lett. \textbf{107}, 069901 (2011)]
doi:10.1103/PhysRevLett.102.161101
[arXiv:0812.4339 [astro-ph]].

\bibitem{Domcke:2017fix}
V.~Domcke, F.~Muia, M.~Pieroni and L.~T.~Witkowski,
``PBH dark matter from axion inflation,''
JCAP \textbf{07}, 048 (2017)
doi:10.1088/1475-7516/2017/07/048
[arXiv:1704.03464 [astro-ph.CO]].

\bibitem{Garcia-Bellido:2017aan}
J.~Garcia-Bellido, M.~Peloso and C.~\"Unal,
``Gravitational Wave signatures of inflationary models from Primordial Black Hole Dark Matter,''
JCAP \textbf{09}, 013 (2017)
doi:10.1088/1475-7516/2017/09/013
[arXiv:1707.02441 [astro-ph.CO]].


\bibitem{Bartolo:2018rku}
N.~Bartolo, V.~De Luca, G.~Franciolini, M.~Peloso, D.~Racco and A.~Riotto,
``Testing primordial black holes as dark matter with LISA,''
Phys. Rev. D \textbf{99}, no.10, 103521 (2019)
doi:10.1103/PhysRevD.99.103521
[arXiv:1810.12224 [astro-ph.CO]].

\bibitem{Cai:2018dig}
R.~g.~Cai, S.~Pi and M.~Sasaki,
``Gravitational Waves Induced by non-Gaussian Scalar Perturbations,''
Phys. Rev. Lett. \textbf{122}, no.20, 201101 (2019)
doi:10.1103/PhysRevLett.122.201101
[arXiv:1810.11000 [astro-ph.CO]].

\bibitem{Unal:2018yaa}
C.~\"Unal,
``Imprints of Primordial Non-Gaussianity on Gravitational Wave Spectrum,''
Phys. Rev. D \textbf{99}, no.4, 041301 (2019)
doi:10.1103/PhysRevD.99.041301
[arXiv:1811.09151 [astro-ph.CO]].


\bibitem{Bird:2016dcv}
S.~Bird, I.~Cholis, J.~B.~Mu\~noz, Y.~Ali-Ha\"\i{}moud, M.~Kamionkowski, E.~D.~Kovetz, A.~Raccanelli and A.~G.~Riess,
``Did LIGO detect dark matter?,''
Phys. Rev. Lett. \textbf{116}, no.20, 201301 (2016)
doi:10.1103/PhysRevLett.116.201301
[arXiv:1603.00464 [astro-ph.CO]].


\bibitem{Clesse:2016vqa}
S.~Clesse and J.~Garc\'\i{}a-Bellido,
``The clustering of massive Primordial Black Holes as Dark Matter: measuring their mass distribution with Advanced LIGO,''
Phys. Dark Univ. \textbf{15}, 142-147 (2017)
doi:10.1016/j.dark.2016.10.002
[arXiv:1603.05234 [astro-ph.CO]].


\bibitem{Niikura:2019kqi}
H.~Niikura, M.~Takada, S.~Yokoyama, T.~Sumi and S.~Masaki,
Phys. Rev. D \textbf{99}, no.8, 083503 (2019)
doi:10.1103/PhysRevD.99.083503
[arXiv:1901.07120 [astro-ph.CO]].

\bibitem{SRcond} 
  A.~Arvanitaki, M.~Baryakhtar and X.~Huang,
 ``Discovering the QCD Axion with Black Holes and Gravitational Waves,''
  Phys.\ Rev.\ D {\bf 91}, no. 8, 084011 (2015)
  doi:10.1103/PhysRevD.91.084011
  [arXiv:1411.2263 [hep-ph]].
  
  
\bibitem{SRcond1} 
  A.~Arvanitaki, M.~Baryakhtar, S.~Dimopoulos, S.~Dubovsky and R.~Lasenby,
  ``Black Hole Mergers and the QCD Axion at Advanced LIGO,''
  Phys.\ Rev.\ D {\bf 95}, no. 4, 043001 (2017)
  doi:10.1103/PhysRevD.95.043001
  [arXiv:1604.03958 [hep-ph]].
  
  

\bibitem{SRreview} 
  R.~Brito, V.~Cardoso and P.~Pani,
  ``Superradiance : Energy Extraction, Black-Hole Bombs and Implications for Astrophysics and Particle Physics,''
  Lect.\ Notes Phys.\  {\bf 906}, pp.1 (2015)
  doi:10.1007/978-3-319-19000-6
  [arXiv:1501.06570 [gr-qc]].
 





\bibitem{Stott:2020gjj}
M.~J.~Stott,
``Ultralight Bosonic Field Mass Bounds from Astrophysical Black Hole Spin,''
[arXiv:2009.07206 [hep-ph]].


\bibitem{Unal:2020jiy}
C.~\"Unal, F.~Pacucci and A.~Loeb,
``Properties of ultralight bosons from heavy quasar spins via superradiance,''
JCAP \textbf{05}, 007 (2021)
doi:10.1088/1475-7516/2021/05/007
[arXiv:2012.12790 [hep-ph]].

\bibitem{Roy:2021uye}
R.~Roy, S.~Vagnozzi and L.~Visinelli,
``Superradiance evolution of black hole shadows revisited,''
Phys. Rev. D \textbf{105}, no.8, 083002 (2022)
doi:10.1103/PhysRevD.105.083002
[arXiv:2112.06932 [astro-ph.HE]].


\bibitem{Chen:2022nbb}
Y.~Chen, R.~Roy, S.~Vagnozzi and L.~Visinelli,
``Superradiant evolution of the shadow and photon ring of Sgr A\ensuremath{\star},''
Phys. Rev. D \textbf{106}, no.4, 043021 (2022)
doi:10.1103/PhysRevD.106.043021
[arXiv:2205.06238 [astro-ph.HE]].

\bibitem{Chen:2021lvo}
Y.~Chen, Y.~Liu, R.~S.~Lu, Y.~Mizuno, J.~Shu, X.~Xue, Q.~Yuan and Y.~Zhao,
``Stringent axion constraints with Event Horizon Telescope polarimetric measurements of M87$^{*}$,''
Nature Astron. \textbf{6}, no.5, 592-598 (2022)
doi:10.1038/s41550-022-01620-3
[arXiv:2105.04572 [hep-ph]].


\bibitem{Pani:2013hpa}
P.~Pani and A.~Loeb,
``Constraining Primordial Black-Hole Bombs through Spectral Distortions of the Cosmic Microwave Background,''
Phys. Rev. D \textbf{88}, 041301 (2013)
doi:10.1103/PhysRevD.88.041301
[arXiv:1307.5176 [astro-ph.CO]].


\bibitem{Baryakhtar:2017ngi}
M.~Baryakhtar, R.~Lasenby and M.~Teo,
``Black Hole Superradiance Signatures of Ultralight Vectors,''
Phys. Rev. D \textbf{96}, no.3, 035019 (2017)
doi:10.1103/PhysRevD.96.035019
[arXiv:1704.05081 [hep-ph]].

\bibitem{Conlon:2017hhi}
J.~P.~Conlon and C.~A.~R.~Herdeiro,
``Can black hole superradiance be induced by galactic plasmas?,''
Phys. Lett. B \textbf{780}, 169-173 (2018)
doi:10.1016/j.physletb.2018.02.073
[arXiv:1701.02034 [astro-ph.HE]].

\bibitem{Dima:2020rzg}
A.~Dima and E.~Barausse,
``Numerical investigation of plasma-driven superradiant instabilities,''
Class. Quant. Grav. \textbf{37}, no.17, 175006 (2020)
doi:10.1088/1361-6382/ab9ce0
[arXiv:2001.11484 [gr-qc]].

\bibitem{Cannizzaro:2020uap}
E.~Cannizzaro, A.~Caputo, L.~Sberna and P.~Pani,
``Plasma-photon interaction in curved spacetime I: formalism and quasibound states around nonspinning black holes,''
Phys. Rev. D \textbf{103}, 124018 (2021)
doi:10.1103/PhysRevD.103.124018
[arXiv:2012.05114 [gr-qc]].


\bibitem{Cannizzaro:2021zbp}
E.~Cannizzaro, A.~Caputo, L.~Sberna and P.~Pani,
``Plasma-photon interaction in curved spacetime. II. Collisions, thermal corrections, and superradiant instabilities,''
Phys. Rev. D \textbf{104}, no.10, 104048 (2021)
doi:10.1103/PhysRevD.104.104048
[arXiv:2107.01174 [gr-qc]].



%
%
%
%
%
%
%
%
%
%
%
%
%
%
%
%
%
  
%
%

%
%
%


  
\bibitem{Flores:2021tmc}
M.~M.~Flores and A.~Kusenko,
``Spins of primordial black holes formed in different cosmological scenarios,''
Phys. Rev. D \textbf{104}, no.6, 063008 (2021)
doi:10.1103/PhysRevD.104.063008
[arXiv:2106.03237 [astro-ph.CO]].
  
  
  
  


\bibitem{Ferreira:2020fam}
E.~G.~M.~Ferreira,
``Ultra-light dark matter,''
Astron. Astrophys. Rev. \textbf{29}, no.1, 7 (2021)
[arXiv:2005.03254 [astro-ph.CO]].


\bibitem{Hlozek:2014lca}
R.~Hlozek, D.~Grin, D.~J.~E.~Marsh and P.~G.~Ferreira,
``A search for ultralight axions using precision cosmological data,''
Phys. Rev. D \textbf{91}, no.10, 103512 (2015)
[arXiv:1410.2896 [astro-ph.CO]].

\bibitem{Poulin:2018dzj}
V.~Poulin, T.~L.~Smith, D.~Grin, T.~Karwal and M.~Kamionkowski,
``Cosmological implications of ultralight axionlike fields,''
Phys. Rev. D \textbf{98}, no.8, 083525 (2018)
[arXiv:1806.10608 [astro-ph.CO]].

\bibitem{Lague:2021frh}
A.~Lagu\"e, J.~R.~Bond, R.~Hlo\v{z}ek, K.~K.~Rogers, D.~J.~E.~Marsh and D.~Grin,
``Constraining ultralight axions with galaxy surveys,''
JCAP \textbf{01}, no.01, 049 (2022)
[arXiv:2104.07802 [astro-ph.CO]].

\bibitem{Bozek:2014uqa}
B.~Bozek, D.~J.~E.~Marsh, J.~Silk and R.~F.~G.~Wyse,
``Galaxy UV-luminosity function and reionization constraints on axion dark matter,''
Mon. Not. Roy. Astron. Soc. \textbf{450}, no.1, 209-222 (2015)
[arXiv:1409.3544 [astro-ph.CO]].

\bibitem{Kobayashi:2017jcf}
T.~Kobayashi, R.~Murgia, A.~De Simone, V.~Ir\v{s}i\v{c} and M.~Viel,
``Lyman-$\alpha$ constraints on ultralight scalar dark matter: Implications for the early and late universe,''
Phys. Rev. D \textbf{96}, no.12, 123514 (2017)
[arXiv:1708.00015 [astro-ph.CO]].

\bibitem{Irsic:2017yje}
V.~Ir\v{s}i\v{c}, M.~Viel, M.~G.~Haehnelt, J.~S.~Bolton and G.~D.~Becker,
``First constraints on fuzzy dark matter from Lyman-$\alpha$ forest data and hydrodynamical simulations,''
Phys. Rev. Lett. \textbf{119}, no.3, 031302 (2017)
[arXiv:1703.04683 [astro-ph.CO]].

\bibitem{Armengaud:2017nkf}
E.~Armengaud, N.~Palanque-Delabrouille, C.~Y\`eche, D.~J.~E.~Marsh and J.~Baur,
``Constraining the mass of light bosonic dark matter using SDSS Lyman-$\alpha$ forest,''
Mon. Not. Roy. Astron. Soc. \textbf{471}, no.4, 4606-4614 (2017)
[arXiv:1703.09126 [astro-ph.CO]].


\bibitem{Rogers:2020ltq}
K.~K.~Rogers and H.~V.~Peiris,
``Strong Bound on Canonical Ultralight Axion Dark Matter from the Lyman-Alpha Forest,''
Phys. Rev. Lett. \textbf{126}, no.7, 071302 (2021)
[arXiv:2007.12705 [astro-ph.CO]].

\bibitem{Rogers:2020cup}
K.~K.~Rogers and H.~V.~Peiris,
``General framework for cosmological dark matter bounds using $N$-body simulations,''
Phys. Rev. D \textbf{103}, no.4, 043526 (2021)
[arXiv:2007.13751 [astro-ph.CO]].


\bibitem{Marsh:2018zyw}
D.~J.~E.~Marsh and J.~C.~Niemeyer,
``Strong Constraints on Fuzzy Dark Matter from Ultrafaint Dwarf Galaxy Eridanus II,''
Phys. Rev. Lett. \textbf{123}, no.5, 051103 (2019)
[arXiv:1810.08543 [astro-ph.CO]].

\bibitem{Bar:2021kti}
N.~Bar, K.~Blum and C.~Sun,
``Galactic rotation curves versus ultralight dark matter: A systematic comparison with SPARC data,''
Phys. Rev. D \textbf{105}, no.8, 8 (2022)
[arXiv:2111.03070 [hep-ph]].


\bibitem{Unal:2022ooa}
C.~Unal, F.~R.~Urban and E.~D.~Kovetz,
``Probing ultralight scalar, vector and tensor dark matter with pulsar timing arrays,''
Phys. Lett. B \textbf{855}, 138830 (2024)
doi:10.1016/j.physletb.2024.138830
[arXiv:2209.02741 [astro-ph.CO]].



\bibitem{Tsukada:2018mbp}
L.~Tsukada, T.~Callister, A.~Matas and P.~Meyers,
``First search for a stochastic gravitational-wave background from ultralight bosons,''
Phys. Rev. D \textbf{99}, no.10, 103015 (2019)
doi:10.1103/PhysRevD.99.103015
[arXiv:1812.09622 [astro-ph.HE]].

\bibitem{Tsukada:2020lgt}
L.~Tsukada, R.~Brito, W.~E.~East and N.~Siemonsen,
``Modeling and searching for a stochastic gravitational-wave background from ultralight vector bosons,''
Phys. Rev. D \textbf{103}, no.8, 083005 (2021)
doi:10.1103/PhysRevD.103.083005
[arXiv:2011.06995 [astro-ph.HE]].



\bibitem{Yuan:2022bem}
C.~Yuan, Y.~Jiang and Q.~G.~Huang,
``Constraints on an ultralight scalar boson from Advanced LIGO and Advanced Virgo\textquoteright{}s first three observing runs using the stochastic gravitational-wave background,''
Phys. Rev. D \textbf{106}, no.2, 023020 (2022)
doi:10.1103/PhysRevD.106.023020
[arXiv:2204.03482 [astro-ph.CO]].

\bibitem{Blas:2019hxz}
D.~Blas, D.~L\'opez Nacir and S.~Sibiryakov,
``Secular effects of ultralight dark matter on binary pulsars,''
Phys. Rev. D \textbf{101} (2020) no.6, 063016
[arXiv:1910.08544 [gr-qc]].


\bibitem{Armaleo:2019gil}
J.~M.~Armaleo, D.~L\'opez Nacir and F.~R.~Urban,
``Binary pulsars as probes for spin-2 ultralight dark matter,''
JCAP \textbf{01} (2020), 053
[arXiv:1909.13814 [astro-ph.HE]].

\bibitem{LopezNacir:2018epg}
D.~L\'opez Nacir and F.~R.~Urban,
``Vector Fuzzy Dark Matter, Fifth Forces, and Binary Pulsars,''
JCAP \textbf{10} (2018), 044
[arXiv:1807.10491 [astro-ph.CO]].

\bibitem{Blas:2016ddr}
D.~Blas, D.~L.~Nacir and S.~Sibiryakov,
``Ultralight Dark Matter Resonates with Binary Pulsars,''
Phys. Rev. Lett. \textbf{118} (2017) no.26, 261102
[arXiv:1612.06789 [hep-ph]].



\bibitem{Mirbabayi:2019uph}
M.~Mirbabayi, A.~Gruzinov and J.~Nore\~na,
``Spin of Primordial Black Holes,''
JCAP \textbf{03}, 017 (2020)
doi:10.1088/1475-7516/2020/03/017
[arXiv:1901.05963 [astro-ph.CO]].

\bibitem{DeLuca:2019buf}
V.~De Luca, V.~Desjacques, G.~Franciolini, A.~Malhotra and A.~Riotto,
``The initial spin probability distribution of primordial black holes,''
JCAP \textbf{05}, 018 (2019)
doi:10.1088/1475-7516/2019/05/018
[arXiv:1903.01179 [astro-ph.CO]].


\bibitem{Harada:2017fjm}
T.~Harada, C.~M.~Yoo, K.~Kohri and K.~I.~Nakao,
``Spins of primordial black holes formed in the matter-dominated phase of the Universe,''
Phys. Rev. D \textbf{96}, no.8, 083517 (2017)
[erratum: Phys. Rev. D \textbf{99}, no.6, 069904 (2019)]
doi:10.1103/PhysRevD.96.083517
[arXiv:1707.03595 [gr-qc]].

\bibitem{DeLuca:2020bjf}
V.~De Luca, G.~Franciolini, P.~Pani and A.~Riotto,
``The evolution of primordial black holes and their final observable spins,''
JCAP \textbf{04}, 052 (2020)
doi:10.1088/1475-7516/2020/04/052
[arXiv:2003.02778 [astro-ph.CO]].


\bibitem{Baryakhtar:2020gao}
M.~Baryakhtar, M.~Galanis, R.~Lasenby and O.~Simon,
``Black hole superradiance of self-interacting scalar fields,''
Phys. Rev. D \textbf{103}, no.9, 095019 (2021)
doi:10.1103/PhysRevD.103.095019
[arXiv:2011.11646 [hep-ph]].

\bibitem{Blas:2020kaa}
D.~Blas and S.~J.~Witte,
``Quenching Mechanisms of Photon Superradiance,''
Phys. Rev. D \textbf{102}, no.12, 123018 (2020)
doi:10.1103/PhysRevD.102.123018
[arXiv:2009.10075 [hep-ph]].


\bibitem{Penrose:1969pc}
R.~Penrose,
``Gravitational collapse: The role of general relativity,''
Riv. Nuovo Cim. \textbf{1}, 252-276 (1969)
doi:10.1023/A:1016578408204



  
%
%
%
%
%
%
%
%
%
%
%
%
%
%
%
%
%
%
%
%
%
%
%
%
%
%
%
%
%
%
%
%
%
%
%
%
%
%
%
%
%
%
%
%
%
%
%
%
%
%
%
%
%
%
%
%
%
%
%
%
%
%
%
%
%
%
%
%
%
%
%
%
%
%
%
%
%
%
%
%
%
%
%
%
%
%
%
%
%
%
%
%
%
%
%
%
%
%
%
%
%
%
%
%
%
%
%
%
%
%
%
%
%
%
%
%
%
%
%



 
\bibitem{Teukolsky:1974yv}
S.~A.~Teukolsky and W.~H.~Press,
``Perturbations of a rotating black hole. III - Interaction of the hole with gravitational and electromagnet ic radiation,''
Astrophys. J. \textbf{193}, 443-461 (1974)
doi:10.1086/153180


\bibitem{Misner:1972kx}
C.~W.~Misner,
``Interpretation of gravitational-wave observations,''
Phys. Rev. Lett. \textbf{28}, 994-997 (1972)
doi:10.1103/PhysRevLett.28.994

\bibitem{Starobinsky:1973aij}
A.~A.~Starobinsky,
``Amplification of waves reflected from a rotating ''black hole''.,''
Sov. Phys. JETP \textbf{37}, no.1, 28-32 (1973)


\bibitem{ZelDovich1972}
Y. B. ZelDovich,
Sov. Phys. JETP \textbf{35}, no.6 ,1085 (1972)

\bibitem{Ternov:1978gq}
I.~M.~Ternov, V.~R.~Khalilov, G.~A.~Chizhov and A.~B.~Gaina,
Sov. Phys. J. \textbf{21}, 1200-1204 (1978)
doi:10.1007/BF00894575

\bibitem{Detweiler:1980uk}
S.~L.~Detweiler,
``KLEIN-GORDON EQUATION AND ROTATING BLACK HOLES,''
Phys. Rev. D \textbf{22}, 2323-2326 (1980)
doi:10.1103/PhysRevD.22.2323


  

  \bibitem{refnovikovthorne}
I.~D.~Novikov and K.~S.~Thorne,
``Astrophysics and black holes''


\bibitem{Brito:2020lup}
R.~Brito, S.~Grillo and P.~Pani,
``Black Hole Superradiant Instability from Ultralight Spin-2 Fields,''
Phys. Rev. Lett. \textbf{124}, no.21, 211101 (2020)
doi:10.1103/PhysRevLett.124.211101
[arXiv:2002.04055 [gr-qc]].

\bibitem{Dias:2023ynv}
O.~J.~C.~Dias, G.~Lingetti, P.~Pani and J.~E.~Santos,
``Black hole superradiant instability for massive spin-2 fields,''
Phys. Rev. D \textbf{108}, no.4, L041502 (2023)
doi:10.1103/PhysRevD.108.L041502
[arXiv:2304.01265 [gr-qc]].
  

\bibitem{Brito:2014wla}
R.~Brito, V.~Cardoso and P.~Pani,
``Black holes as particle detectors: evolution of superradiant instabilities,''
Class. Quant. Grav. \textbf{32}, no.13, 134001 (2015)
doi:10.1088/0264-9381/32/13/134001
[arXiv:1411.0686 [gr-qc]].
  
  
 
 
\bibitem{SRconstraints} 
  M.~J.~Stott and D.~J.~E.~Marsh,
  ``Black hole spin constraints on the mass spectrum and number of axionlike fields,''
  Phys.\ Rev.\ D {\bf 98}, no. 8, 083006 (2018)
  doi:10.1103/PhysRevD.98.083006
  [arXiv:1805.02016 [hep-ph]].
  



  
    
\bibitem{SRnumanal} 
  H.~Witek, V.~Cardoso, A.~Ishibashi and U.~Sperhake,
  ``Superradiant instabilities in astrophysical systems,''
  Phys.\ Rev.\ D {\bf 87}, no. 4, 043513 (2013)
  doi:10.1103/PhysRevD.87.043513
  [arXiv:1212.0551 [gr-qc]].

%
%
%
%
%
%
%
%
%
%
%
%
%
%
%
%
%
%
%
%
%
%
%


\bibitem{Hawking:1974rv}
S.~W.~Hawking,
``Black hole explosions,''
Nature \textbf{248}, 30-31 (1974)
doi:10.1038/248030a0


\bibitem{Hawking:1975vcx}
S.~W.~Hawking,
``Particle Creation by Black Holes,''
Commun. Math. Phys. \textbf{43}, 199-220 (1975)
[erratum: Commun. Math. Phys. \textbf{46}, 206 (1976)]
doi:10.1007/BF02345020


\bibitem{Page:1976df}
D.~N.~Page,
``Particle Emission Rates from a Black Hole: Massless Particles from an Uncharged, Nonrotating Hole,''
Phys. Rev. D \textbf{13}, 198-206 (1976)
doi:10.1103/PhysRevD.13.198

\bibitem{Page:1976ki}
D.~N.~Page,
``Particle Emission Rates from a Black Hole. 2. Massless Particles from a Rotating Hole,''
Phys. Rev. D \textbf{14}, 3260-3273 (1976)
doi:10.1103/PhysRevD.14.3260

\bibitem{Page:1977um}
D.~N.~Page,
``Particle Emission Rates from a Black Hole. 3. Charged Leptons from a Nonrotating Hole,''
Phys. Rev. D \textbf{16}, 2402-2411 (1977)
doi:10.1103/PhysRevD.16.2402


\bibitem{MacGibbon:1990zk}
J.~H.~MacGibbon and B.~R.~Webber,
Phys. Rev. D \textbf{41}, 3052-3079 (1990)
doi:10.1103/PhysRevD.41.3052


\bibitem{Khlopov:1985fch}
M.~Y.~Khlopov, B.~A.~Malomed, I.~B.~Zeldovich and Y.~B.~Zeldovich,
Mon. Not. Roy. Astron. Soc. \textbf{215}, no.4, 575-589 (1985)
doi:10.1093/mnras/215.4.575

\bibitem{Rosa:2017ury}
J.~G.~Rosa and T.~W.~Kephart,
``Stimulated Axion Decay in Superradiant Clouds around Primordial Black Holes,''
Phys. Rev. Lett. \textbf{120}, no.23, 231102 (2018)
doi:10.1103/PhysRevLett.120.231102
[arXiv:1709.06581 [gr-qc]].

\bibitem{Bernal:2021yyb}
N.~Bernal, F.~Hajkarim and Y.~Xu,
``Axion Dark Matter in the Time of Primordial Black Holes,''
Phys. Rev. D \textbf{104}, 075007 (2021)
doi:10.1103/PhysRevD.104.075007
[arXiv:2107.13575 [hep-ph]].

\bibitem{Branco:2023frw}
N.~P.~Branco, R.~Z.~Ferreira and J.~G.~Rosa,
``Superradiant axion clouds around asteroid-mass primordial black holes,''
[arXiv:2301.01780 [hep-ph]].

\bibitem{Calza:2023rjt}
M.~Calz\`a, J.~G.~Rosa and F.~Serrano,
JHEP \textbf{05}, 140 (2024)
doi:10.1007/JHEP05(2024)140
[arXiv:2306.09430 [hep-ph]].


\bibitem{Richarte:2021fbi}
M.~G.~Richarte, \'E.~L.~Martins and J.~C.~Fabris,
``Scattering and absorption of a scalar field impinging on a charged black hole in the Einstein-Maxwell-dilaton theory,''
Phys. Rev. D \textbf{105}, no.6, 064043 (2022)
doi:10.1103/PhysRevD.105.064043
[arXiv:2111.01595 [gr-qc]].

\bibitem{Mazde:2022sdx}
K.~Mazde and L.~Visinelli,
``The interplay between the dark matter axion and primordial black holes,''
JCAP \textbf{01}, 021 (2023)
doi:10.1088/1475-7516/2023/01/021
[arXiv:2209.14307 [astro-ph.CO]].


\bibitem{Hooper:2019gtx}
D.~Hooper, G.~Krnjaic and S.~D.~McDermott,
``Dark Radiation and Superheavy Dark Matter from Black Hole Domination,''
JHEP \textbf{08}, 001 (2019)
doi:10.1007/JHEP08(2019)001
[arXiv:1905.01301 [hep-ph]].


\bibitem{Papanikolaou:2020qtd}
T.~Papanikolaou, V.~Vennin and D.~Langlois,
``Gravitational waves from a universe filled with primordial black holes,''
JCAP \textbf{03}, 053 (2021)
doi:10.1088/1475-7516/2021/03/053
[arXiv:2010.11573 [astro-ph.CO]].




\bibitem{Bernal:2022oha}
N.~Bernal, Y.~F.~Perez-Gonzalez and Y.~Xu,
``Superradiant production of heavy dark matter from primordial black holes,''
Phys. Rev. D \textbf{106}, no.1, 015020 (2022)
doi:10.1103/PhysRevD.106.015020
[arXiv:2205.11522 [hep-ph]].


\bibitem{March-Russell:2022zll}
J.~March-Russell and J.~G.~Rosa,
``Micro-Bose/Proca dark matter stars from black hole superradiance,''
[arXiv:2205.15277 [gr-qc]].


\bibitem{Gehrman:2022imk}
T.~C.~Gehrman, B.~Shams Es Haghi, K.~Sinha and T.~Xu,
``Baryogenesis, Primordial Black Holes and MHz-GHz Gravitational Waves,''
[arXiv:2211.08431 [hep-ph]].




\bibitem{Franciolini:2022pav}
G.~Franciolini and A.~Urbano,
``Primordial black hole dark matter from inflation: The reverse engineering approach,''
Phys. Rev. D \textbf{106}, no.12, 123519 (2022)
doi:10.1103/PhysRevD.106.123519
[arXiv:2207.10056 [astro-ph.CO]].

\bibitem{Chakraborty:2022mwu}
A.~Chakraborty, P.~K.~Chanda, K.~L.~Pandey and S.~Das,
``Formation and Abundance of Late-forming Primordial Black Holes as Dark Matter,''
Astrophys. J. \textbf{932}, no.2, 119 (2022)
doi:10.3847/1538-4357/ac6ddd
[arXiv:2204.09628 [astro-ph.CO]].

\bibitem{nosuperradiancebyinteractions1}
H.~Fukuda and K.~Nakayama,
``Aspects of Nonlinear Effect on Black Hole Superradiance,''
JHEP \textbf{01}, 128 (2020)
doi:10.1007/JHEP01(2020)128
[arXiv:1910.06308 [hep-ph]].
  
  
\bibitem{nosuperradiancebyinteractions2}
A.~Mathur, S.~Rajendran and E.~H.~Tanin,
``Clockwork mechanism to remove superradiance limits,''
Phys. Rev. D \textbf{102}, no.5, 055015 (2020)
doi:10.1103/PhysRevD.102.055015
[arXiv:2004.12326 [hep-ph]].

\bibitem{Cannizzaro:2022xyw}
E.~Cannizzaro, L.~Sberna, A.~Caputo and P.~Pani,
``Dark photon superradiance quenched by dark matter,''
Phys. Rev. D \textbf{106}, no.8, 083019 (2022)
doi:10.1103/PhysRevD.106.083019
[arXiv:2206.12367 [hep-ph]].



\bibitem{Yoshino:2012kn}
H.~Yoshino and H.~Kodama,
``Bosenova collapse of axion cloud around a rotating black hole,''
Prog. Theor. Phys. \textbf{128}, 153-190 (2012)
doi:10.1143/PTP.128.153
[arXiv:1203.5070 [gr-qc]].

  


\bibitem{spin2SR}  
  R.~Brito, V.~Cardoso and P.~Pani,
  ``Massive spin-2 fields on black hole spacetimes: Instability of the Schwarzschild and Kerr solutions and bounds on the graviton mass,''
  Phys.\ Rev.\ D {\bf 88} (2013) no.2,  023514
  doi:10.1103/PhysRevD.88.023514
  [arXiv:1304.6725 [gr-qc]].




\bibitem{Babichev:2016bxi}
E.~Babichev, L.~Marzola, M.~Raidal, A.~Schmidt-May, F.~Urban, H.~Veerm\"ae and M.~von Strauss,
``Heavy spin-2 Dark Matter,''
JCAP \textbf{09}, 016 (2016)
doi:10.1088/1475-7516/2016/09/016
[arXiv:1607.03497 [hep-th]].


\bibitem{Poddar:2019zoe}
T.~Kumar Poddar, S.~Mohanty and S.~Jana,
``Constraints on ultralight axions from compact binary systems,''
Phys. Rev. D \textbf{101}, no.8, 083007 (2020)
doi:10.1103/PhysRevD.101.083007
[arXiv:1906.00666 [hep-ph]].




\bibitem{Ferraz:2020zgi}
P.~B.~Ferraz, T.~W.~Kephart and J.~G.~Rosa,
``Superradiant pion clouds around primordial black holes,''
JCAP \textbf{07}, no.07, 026 (2022)
doi:10.1088/1475-7516/2022/07/026
[arXiv:2004.11303 [gr-qc]].


\bibitem{ParticleDataGroup:2022pth}
R.~L.~Workman \textit{et al.} [Particle Data Group],
``Review of Particle Physics,''
PTEP \textbf{2022}, 083C01 (2022)
doi:10.1093/ptep/ptac097


  
  
\bibitem{Hu:2000ke}
W.~Hu, R.~Barkana and A.~Gruzinov,
``Cold and fuzzy dark matter,''
Phys. Rev. Lett. \textbf{85}, 1158-1161 (2000)
doi:10.1103/PhysRevLett.85.1158
[arXiv:astro-ph/0003365 [astro-ph]].
%
%
\bibitem{Hui:2016ltb}
L.~Hui, J.~P.~Ostriker, S.~Tremaine and E.~Witten,
``Ultralight scalars as cosmological dark matter,''
Phys. Rev. D \textbf{95}, no.4, 043541 (2017)
doi:10.1103/PhysRevD.95.043541
[arXiv:1610.08297 [astro-ph.CO]].
%
%
%
\bibitem{Marsh:2015xka}
D.~J.~E.~Marsh,
``Axion Cosmology,''
Phys. Rept. \textbf{643}, 1-79 (2016)
doi:10.1016/j.physrep.2016.06.005
[arXiv:1510.07633 [astro-ph.CO]].
%




\bibitem{Marzola:2017lbt}
L.~Marzola, M.~Raidal and F.~R.~Urban,
``Oscillating Spin-2 Dark Matter,''
Phys. Rev. D \textbf{97}, no.2, 024010 (2018)
doi:10.1103/PhysRevD.97.024010
[arXiv:1708.04253 [hep-ph]].










%
%
%
%
%
%
%
%
%
%
%
%
%
%
%
%
%
%
%
%
%
%
%
%
%
%
%
%
%
%
%
%
%
%
%
%
%
%
%
%
%
%
%
%
%
%
%
%
%
%
%
%
%
%
%
%
%
%
%
%
%
%
%
%
%
%
%
%
%
%
%
%
%
%
%
%
%
%
%
%
%
%
%
%
%
%
%
%
%
%
%
%
%
%
%
%
%
%
%
%
%
%
%
%
%
%
%
%
%
%
%
%
%
%
%
%
%
%
%
%
%
%
%
%
%
%
%

  
    
%
%
%
%
%
%
%
%
%
%
%


  
\end{thebibliography}
\end{document}